\documentclass[aps,reprint,superscriptaddress,amsmath,amssymb,longbibliography]{revtex4-1}

\usepackage{graphicx}
\usepackage{amsmath}
\usepackage{bm}

\usepackage{color}

\begin{document}

%\preprint{}

\title{Surges of Collective Human Activity Emerge From Simple Pairwise Correlations}


\author{Christopher W. Lynn}
\author{Lia Papadopoulos}
\affiliation{Department of Physics and Astronomy, University of Pennsylvania, Philadelphia, PA 19104, USA}
\author{Daniel D. Lee}
\affiliation{Department of Electrical \& Systems Engineering, University of Pennsylvania, Philadelphia, PA 19104, USA}
\author{Danielle S. Bassett}
%\altaffiliation{To whom correspondence should be addressed.}
\affiliation{Department of Physics and Astronomy, University of Pennsylvania, Philadelphia, PA 19104, USA}
\affiliation{Department of Electrical \& Systems Engineering, University of Pennsylvania, Philadelphia, PA 19104, USA}
\affiliation{Department of Bioengineering, University of Pennsylvania, Philadelphia, PA 19104, USA}
\affiliation{Department of Neurology, University of Pennsylvania, Philadelphia, PA 19104, USA}

%\thanks{}


\date{\today}

\begin{abstract}

Human populations exhibit complex behaviors---characterized by long-range correlations and surges in activity---across a range of social, political, and technological contexts. Yet it remains unclear where these collective behaviors come from, or if there even exists a set of unifying principles. Indeed, existing explanations typically rely on context-specific mechanisms, such as traffic jams driven by work schedules or spikes in online traffic induced by significant events. However, analogies with statistical mechanics suggest a more general mechanism: that collective patterns can emerge organically from fine-scale interactions within a population. Here, across four different modes of human activity, we show that the simplest correlations in a population---those between pairs of individuals---can yield accurate quantitative predictions for the large-scale behavior of the entire population. To quantify the minimal consequences of pairwise correlations, we employ the principle of maximum entropy, making our description equivalent to an Ising model whose interactions and external fields are notably calculated from past observations of population activity. In addition to providing accurate quantitative predictions, we show that the topology of learned Ising interactions resembles the network of inter-human communication within a population. Together, these results demonstrate that fine-scale correlations can be used to predict large-scale social behaviors, a perspective that has critical implications for modeling and resource allocation in human populations.

\end{abstract}

%\pacs{asdfasdf}
%\keywords{keywords}

\maketitle

\section{Introduction}

In the study of human behavior, significant effort has focused on understanding the actions of one or two individuals at a time. It has been observed, for instance, that people engage in ``bursts'' of actions in quick succession \cite{Barabasi-01,Vazquez-01,Rybski-01}, and significant effort has concentrated on understanding the correlated activity of pairs and triplets of individuals \cite{Eckmann-01,Rybski-01}. But if we broaden our perspective to an entire population, it becomes increasingly clear that humans also exhibit large-scale patterns of correlated activity. For example, urban transportation systems undergo surges of correlated activity known as traffic jams \cite{Peng-01}, first responders are required to handle correlated spikes in demand for emergency services \cite{Bagrow-01}, and internet and telephone networks must be designed to withstand surges of collective activity \cite{Crane-01,Candia-01}. But where do these large-scale patterns come from? Does it even make sense to discuss such distinct phenomena in the same breath?

Existing explanations for collective human behaviors have focused primarily on external mechanisms, such as fluctuations in urban traffic based on the time of the week \cite{Peng-01} or spikes in demand for emergency services in response to natural disasters \cite{Bagrow-01}. While external influences are an important part of the story, such explanations are inherently limited by their reliance on context-specific mechanisms like daily and weekly rhythms and natural disasters. By contrast, 
interactions between individuals are present in almost every human context, providing the possibility for a much more general explanation for the emergence of large-scale correlations. Precisely this line of reasoning has fostered vibrant efforts linking the study of social systems to tools and intuitions from statistical physics \cite{Castellano-01}. By adapting established models of collective behavior in physical systems, such as the Ising model and similar agent-based models, scientists have gained a deeper understanding of the nature of collective behaviors in social systems. This program, for example, has resulted in Ising-like models of social dynamics and human cooperation \cite{Perc-01,Galam-02,Galam-04}, viral models aiding in the design of vaccination strategies \cite{Wang-01}, descriptions of the evolution of social networks \cite{Barabasi-02}, and statistical models of criminal activity \cite{DOrsogna-01,Helbing-01}.

Here we draw inspiration from these seminal results to investigate the role of fine-scale correlations in generating large-scale patterns of human activity. Focusing on four datasets of human activity, from email and private message correspondence to physical contact and music streaming, we find that each population exhibits periods of intense collective activity, which cannot be explained by commonly-used models that assume independence in human behavior \cite{Haight-01,Rasch-01,Karagiannis-01,Gerlough-01}. Intuitively, these surges in activity could be driven by a common external influence, such as people's daily and weekly schedules. Instead, to quantify the collective impact of pairwise correlations, we construct a pairwise maximum entropy model that is formally equivalent to an Ising model from statistical mechanics. While the Ising model has previously been used to understand qualitative aspects of human activity \cite{Castellano-01,Galam-02,Galam-05,Galam-01,Lynn-01}, here, in order to make quantitative predictions, we calculate the specific external fields and pairwise interactions that best describe each population. In what follows, we show that this maximum entropy model (i) accurately predicts the frequencies of different patterns of collective human activity, and (ii) bears a close resemblance to the network of inter-human communication within a population. Taken together, these results constitute an important step in the development of quantitative models of collective human behavior based on fine-scale correlations within a population. Such models, in turn, have important implications for resource allocation in communication \cite{Candia-01} and transportation \cite{Peng-01} networks, understanding social organization \cite{Onnela-01}, and preventing viral epidemics \cite{Pastor-01}.

\section{The network effects of correlations}

As a salient example of collective human activity, we begin by studying patterns of email correspondence, focusing specifically on the email activity of 100 scientists at a European research institution over 526 days \cite{Paranjape-01, Panzarasa-01}. To understand the role of correlations in the timing of people's actions---and in order to compare against other types of activities that are not directed from one individual to another \cite{Crane-01,Peng-01,Bagrow-01,Sornette-01,Deschatres-01}---we initially focus on the timing of sent emails, while blinding our analysis to the email recipients. Importantly, this will later allow us to compare the architecture of functional interactions derived from our maximum entropy model with the network of communication within the population. 

\begin{figure}
\centering
\includegraphics[width = .49\textwidth]{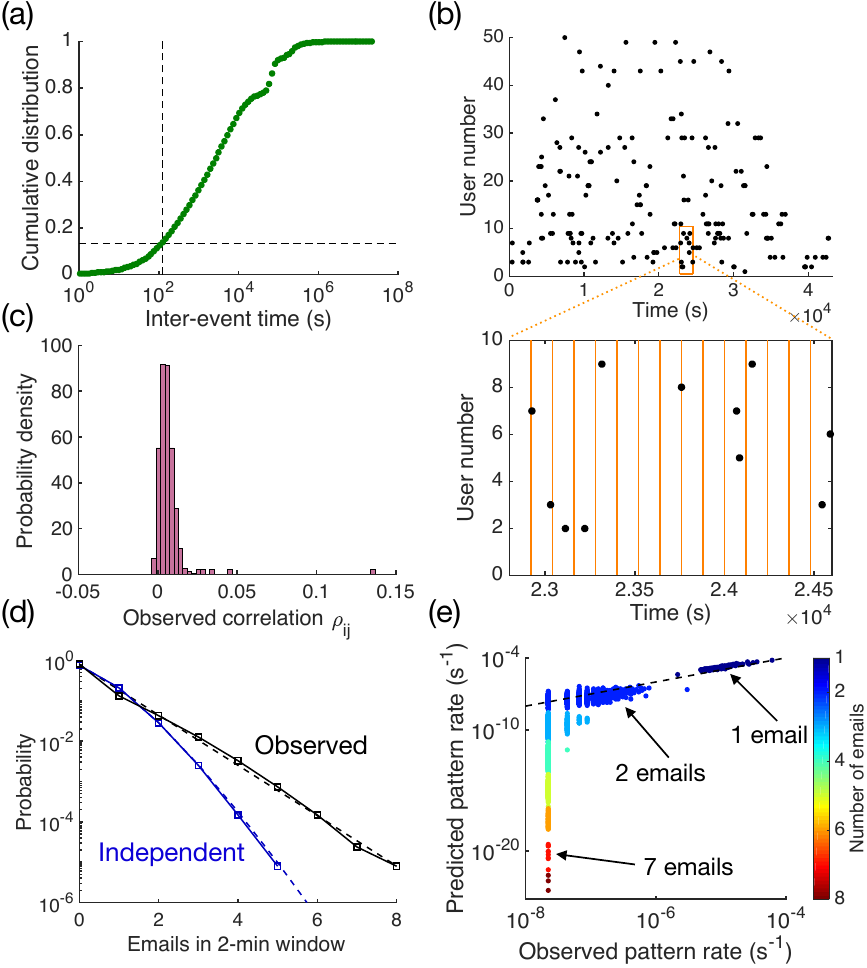}
\caption{\label{correlations} Surges of human activity and failure of the independent approximation. (a) Distribution of inter-event times for individuals in a network of email correspondence. The dashed lines indicate the proportion of inter-event times less than two minutes. (b) Top: Activity of the 50 most active individuals over a half-day period, where each dot represents a sent email. Bottom: Network activity is discretized into two-minute windows. (c) Histogram of Pearson correlation coefficients $\rho_{ij}$ between activity time series for all pairs in the 100-person population. (d) Distribution of the number of emails sent in a given two-minute window (black) and the distribution after shuffling each person's activity to eliminate correlations (blue). The dashed lines show an exponential distribution fit to the observed data (black) and a Poisson distribution fit to the shuffled data (blue). (e) The rate of each observed activity pattern, plotted against the approximate pattern rate assuming independent people. The dashed line indicates equality.}
\end{figure}

In a sufficiently small window of time $\Delta t$, each action appears binary---either individual $i$ sent an email ($\sigma_i = 1$) or they were silent ($\sigma_i = 0$). By discretizing human activity in this way, we can begin to quantify correlations between people's actions. We wish for the time window $\Delta t$ to be as large as possible (to detect correlations between individuals) without being so large that individuals perform multiple actions within the same window. We find that nearly 90\% of consecutive emails from the same individual are sent with at least two minutes in between [Fig. \ref{correlations}(a)], defining a natural time scale that we use as our $\Delta t$. Discretizing the data, as shown in Fig. \ref{correlations}(b), we produce a set of $\sim 3.8\times 10^5$ binary vectors (patterns) $\bm{\sigma}$, each of which captures the activity of the entire population within a given two-minute window.

The simplest and most common models of human activity assume that each individual behaves independently, implying that the number of people performing an action in a given window follows a Poisson distribution \cite{Haight-01}. Indeed, the Poisson distribution has been widely used to quantify the effects of various human actions, including telephone calls to a call center \cite{Rasch-01}, internet activity \cite{Karagiannis-01}, industrial accidents \cite{Haight-01,Rasch-01}, and highway traffic flow \cite{Gerlough-01}. In our population of email users, most pairs of individuals are only weakly correlated [Fig. \ref{correlations}(c)], suggesting that small groups should be well-approximated by an independent model. However, if we extend the independent approximation to the entire population of 100 email users, it fails dramatically. While the Poisson distribution predicts a super-exponential drop off in the number of actions performed in a given window, we find instead that human activity actually follows an exponential distribution [Fig. \ref{correlations}(d)]. This exponential distribution is characterized by a heavy tail, representing moments in time when many more people are sending emails than would be expected if they were behaving independently. Additionally, we report similar heavy-tailed distributions in separate datasets of private messages, physical contacts, and music streams [Figs. \ref{Irvine_fig}-\ref{LastFM_fig}]. For comparison, after shuffling the timing of the emails to eliminate correlations \cite{Schneidman-01}, we do not witness a window involving six or more active users [Fig. \ref{correlations}(d)], while we do observe $\sim$1500 such instances in the original dataset---nearly three per day.

The independent approximation also makes straightforward predictions for the rate of each activity pattern. Denoting the probability of individual $i$ sending an email in a given two-minute window by $p_i(\sigma_i)$, the probability of observing a given activity pattern $\bm{\sigma}$ is simply predicted to be $P_1(\bm{\sigma}) = \prod_i p_i(\sigma_i)$. This independent model severely under-predicts patterns involving three or more active email users [Fig. \ref{correlations}(e)], and we find a similar discrepancy in a network of private messages [Fig. \ref{Irvine_fig}(c)]. In fact, under the independent model, each pattern of email activity involving seven active users should have only appeared roughly once every $10^{20}$ seconds---longer than the age of the universe. We conclude that the independent approximation fails to explain the heavy-tailed nature of human behavior, characterized by surges of collective activity \cite{Candia-01,Crane-01,Peng-01,Bagrow-01}. But where do these surges come from?

\section{A maximum entropy model of human activity}

\begin{figure}
\includegraphics[width = .49\textwidth]{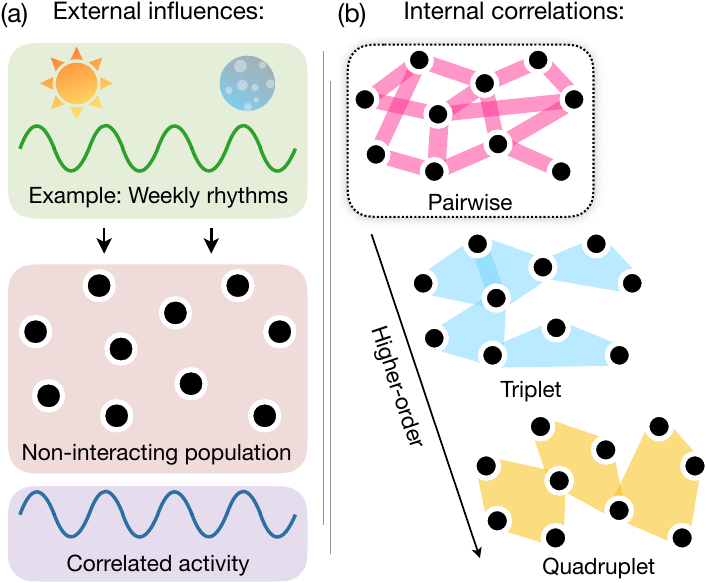}
\caption{\label{Influence} External influences versus internal correlations. (a) An external mechanism---here taken to be weekly rhythms---influencing the activity of a population of non-interacting humans. Intuitively, circadian and weekly rhythms might influence people to send emails more frequently during the daytime and on weekdays, thereby inducing population-wide correlations. (b) Alternatively, population-wide correlations could arise from fine-scale interactions between individuals within a population. The set of all correlations forms a hierarchy, beginning with simple pairwise correlations between two individuals, followed by more complicated higher-order correlations involving three (triplet), four (quadruplet), or more individuals.}
\end{figure}

To improve upon the independent model, we must take into account correlations between individuals. Intuitively, such correlations could be driven by external influences such as daily and weekly rhythms [Fig. \ref{Influence}(a)], a hypothesis that has dominated existing explanations of large-scale human behaviors \cite{Peng-01,Bagrow-01,Crane-01,Candia-01}. Alternatively, fine-scale correlations involving only a few individuals could build upon one another to have a strong impact on the population as a whole [Fig. \ref{Influence}(b)]. Here, we focus on the simplest possible correlations within a population---those between pairs of individuals---and ask whether these pairwise correlations can give rise to the large-scale patterns of activity that we observe in the data. As we will see, focusing on pairwise correlations represents a natural first step towards understanding emergent collective human activity, opening the door for straightforward generalizations to more complex higher-order correlations [Fig. \ref{Influence}(b)] \cite{Ganmor-01, Marre-01}.

We require a model that incorporates the observed pairwise correlations in the data, while including as little information as possible about higher-order correlations between three, four, or more individuals. While it is not immediately obvious how one would construct such a model, Jaynes famously showed that an elegant solution lies in the principle of maximum entropy \cite{Jaynes-01}: Among the infinite set of distributions consistent with a given set of correlations, the unique one that assumes as little information as possible about additional correlations is precisely the distribution with maximum entropy. This maximum entropy principle lies at the heart of equilibrium statistical mechanics \cite{Jaynes-01, Cover-01} and has become increasingly popular as a tool for studying emergent phenomena in a range of complex systems, including networks of neurons in the brain \cite{Schneidman-01,Ganmor-01}, flocks of birds \cite{Bialek-01}, protein structures \cite{Weigt-01}, and gene coexpression patterns \cite{Lezon-01}. Despite this widespread adoption in biophysics, to our knowledge a similar data-driven approach has not previously been attempted in the social sciences.

Here we consider the pairwise maximum entropy model, defined by the Boltzmann distribution
\begin{equation}
\label{P2}
P_2(\bm{\sigma}) = \frac{1}{Z}\exp\Bigg(\sum_i h_i\sigma_i + \frac{1}{2}\sum_{i\neq j} J_{ij}\sigma_i\sigma_j\Bigg) ,
\end{equation}
where the external fields $h_i$ and pairwise interactions $J_{ij}$ are Lagrange multipliers that ensure the model matches the observed individual activity rates and pairwise correlations in the data, respectively, and $Z$ is the normalizing partition function. If we switch notation to $\sigma_i = \pm 1$, where $+1$ stands for activity and $-1$ for inactivity, $P_2$ is equivalent to the Ising model, which has long been used to simulate human dynamics in social networks \cite{Castellano-01,Galam-02,Galam-05,Galam-01,Lynn-01}. However, while existing applications of the Ising model to human populations are based on metaphors about how people interact \cite{Galam-02,Galam-04,Galam-05,Lynn-01,Lynn-02}, we emphasize that our use of the Ising model is quantitatively rigorous in the sense that the external fields $h_i$ and interactions $J_{ij}$ are calculated to fit the observed activity of a given population (see Appendix \ref{BM}).

\section{The minimal consequences of pairwise correlations}

\begin{figure}
\includegraphics[width = .49\textwidth]{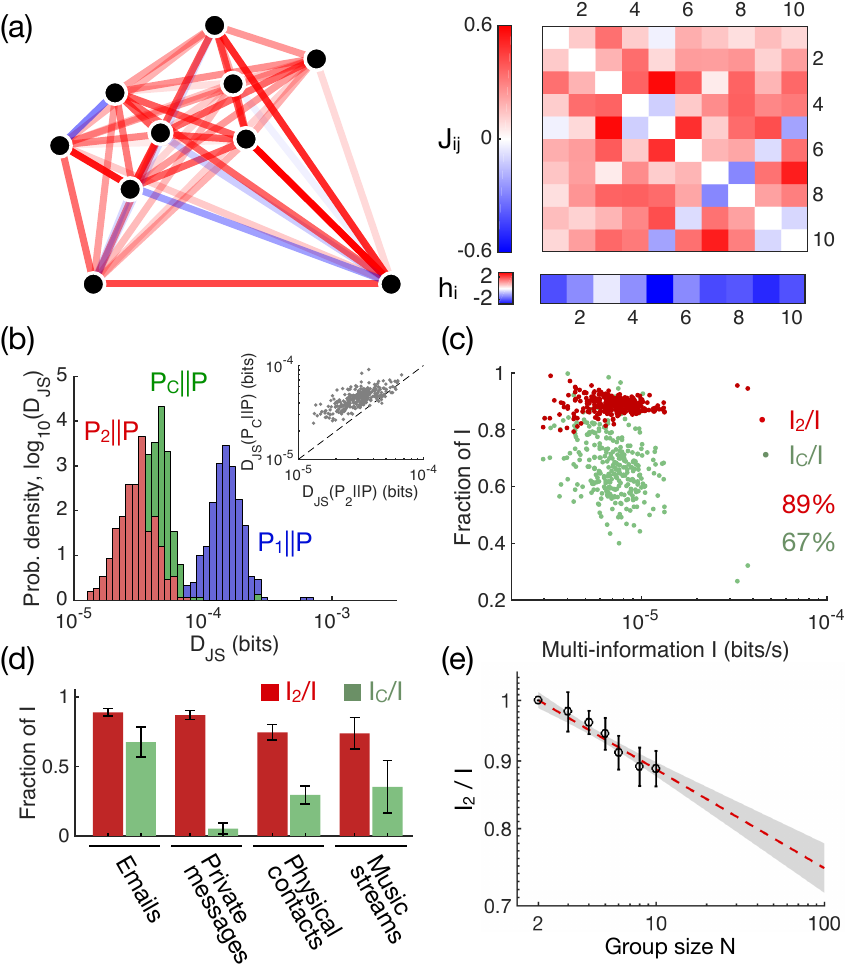}
\caption{\label{Ising} The pairwise maximum entropy model accurately describes human behavior. (a) Learned Ising interactions $J_{ij}$ and external fields $h_i$ describing a random 10-person group in the email network. (b) Jensen-Shannon divergences between the true distribution $P$ and the independent $P_1$ (blue), maximum entropy $P_2$ (red), and conditionally independent $P_C$ (green) models. Histograms reflect estimates from 300 random groups of 10 individuals. Inset: $D_{JS}(P_2||P)$ versus $D_{JS}(P_C||P)$ for the 300 groups. The dashed line indicates equality. (c) Fraction of the network correlation (quantified by the multi-information $I$) captured by the maximum entropy (red) and conditionally independent (green) models, plotted against $I$ for each group of 10 people. The multi-information is divided by $\Delta t$ to remove dependence on the window size. (d) Fraction of the total correlation captured by the pairwise (red) and conditionally independent (green) models in four different modes of human activity: email correspondence, private messaging, physical interactions, and online music streaming. Error bars represent standard deviations over 300 random 10-person groups for the email and private message datasets and over 200 groups for the physical contact and music streaming datasets. (e) Fraction of the multi-information in the email data captured by the maximum entropy model versus group size, where each data point is averaged over 300 randomly-selected groups. The dashed line represents the best log-linear fit, with 95\% confidence interval indicated by the shaded region.}
\end{figure}

Calculations in the Ising model typically require summing over all $2^N$ activity patterns, where $N$ is the number of elements in a system, prohibiting applications to large populations. Thus, it is common to construct a picture of the whole population by studying many different sub-populations \cite{Schneidman-01}, such as the 10 email users in Fig. \ref{Ising}(a). To quantify the explanatory power of pairwise correlations, we need meaningful ways to compare the accuracy of the maximum entropy model $P_2$ to that of the independent model $P_1$. Toward this end, we use the Jensen-Shannon divergence $D_{JS}(Q||P)$ as a measure of distance from each of the model distributions (call them $Q$) to the observed activity distribution $P$. Put simply, the Jensen-Shannon divergence represents the inverse of the number of independent samples needed to distinguish each model $Q$ from the observed data \cite{Lin-01}. Across 300 random groups of 10 users, we find that on average one would require $3.13\times 10^4$ independent samples---over 43 days worth of data---to distinguish the pairwise model $P_2$ from the true distribution $P$ [Fig. \ref{Ising}(b)]. By contrast, one would typically require five times fewer samples to distinguish the independent model $P_1$ from the observed data. Moreover, we find qualitatively similar results for individuals engaged in private messaging [Fig. \ref{Irvine_fig}(e)], face-to-face interactions [Fig. \ref{Conference_fig}(c)], and online music streaming [Fig. \ref{LastFM_fig}(c)]. These observations suggest that the pairwise model provides a marked improvement in accuracy over the independent model.

We also wish to compare against a model representing the hypothesis that patterns of human activity are driven by external influences. While there are many external factors influencing human actions on a daily basis, from weather patterns to shifting demands at work, here we consider the most intuitive and well-studied external influence; namely, the impact of daily and weekly routines [see Fig. \ref{Influence}(a)] \cite{Peng-01,Crane-01,Candia-01,Malmgren-02}. To formalize the hypothesis that activity patterns are driven by daily and weekly schedules, we consider the conditionally independent model $P_C$, wherein each individual performs actions independently from all other individuals, but their activity rates are allowed to vary based on the time of the week \cite{Schneidman-01, Barlow-01} (see Appendix \ref{CondInd}). Compared to the conditionally independent model $P_C$, we find that the maximum entropy model $P_2$ is closer to the observed data (i.e., has a smaller Jensen-Shannon divergence from $P$) across 291 of the 300 groups [Fig. \ref{Ising}(c), Inset]. This result is particularly notable when considering that $P_2$ only has 55 parameters for each group of 10 individuals, while $P_C$ requires knowledge of each individual's email rate at each time during the week, totaling over $5\times 10^4$ parameters.

The pairwise model accurately predicts the rates of particular activity patterns, but does it explain a majority of the total correlation in the population? To answer this question, we note that the total amount of correlation in the network, contributed by correlations between groups of users of all sizes, is quantified by the multi-information $I=S_1 - S$, where $S_1$ is the entropy of the independent distribution $P_1$ and $S$ is the entropy of the observed distribution $P$ \cite{Cover-01} (see Appendix \ref{Entropy}). To determine the amount of multi-information that is contributed by pairwise correlations, it is useful to review the properties of maximum entropy models. For a population of $N$ elements, we can define a sequence of maximum entropy models $P_k$ that are consistent with all correlations up to the $k^{\text{th}}$-order, where $k=1,2,\hdots,N$. These models form a hierarchy, from $P_1$, in which all elements are independent, up to $P_N$, which is an exact description of the observed activity. As we climb up this hierarchy, the entropies $S_k$ of the distributions decrease monotonically toward the true entropy ($S_1\ge S_2\ge\cdots\ge S_N = S$); and the combined contribution of all $k^{\text{th}}$-order correlations is quantified by the entropy difference $I_k = S_{k-1} - S_k$. We note, for instance, that these entropy differences sum to the full multi-information: $I_2 + \cdots + I_N = I$. Thus, the problem of determining how much of the total correlation in the data stems from pairwise correlations formally reduces to calculating the proportion of the multi-information $I$ that is accounted for by the reduction in entropy from pairwise correlations (i.e., $I_2 = S_1 - S_2$).

We observe that pairwise correlations account for a striking $I_2/I\approx 89\%$ of the total correlation in groups of 10 users [Fig. \ref{Ising}(c)]. In turn, this observation implies that the contributions of all other higher-order correlations, $I_3 + \cdots + I_N$, only combine to account for the remaining $11\%$ of the multi-information. Meanwhile, the amount of correlation attributable to daily and weekly rhythms is represented by the entropy difference $I_C = S_1 - S_C$, where $S_C$ is the entropy of the conditionally independent model $P_C$. This popular explanation for collective human behavior is consistently less effective than the maximum entropy model at capturing the correlations in the data [$I_C/I\approx 67\%$; Fig. \ref{Ising}(c)]. Importantly, we show (i) that these results are robust to both reasonable variation in the time window $\Delta t$ used to discretize the data [Appendix \ref{Dt}, Fig. \ref{ChangeDt}] as well as differences in the set of individuals selected for analysis [Appendix \ref{Users}, Fig. \ref{ChangeUsers}], and (ii) that the maximum entropy model is relatively consistent over time [Appendix \ref{Consistency}, Fig. \ref{TimeComp}]. Moreover, we verify that similar results hold in separate datasets of private messages [Appendix \ref{Irvine}, Fig. \ref{Irvine_fig}], physical contacts between individuals [Appendix \ref{Conference}, Fig. \ref{Conference_fig}], and music streaming online [Appendix \ref{LastFM}, Fig. \ref{LastFM_fig}], as summarized in Fig. \ref{Ising}(d). In the dataset of private messages, for instance, the pairwise model captures nearly the same amount of correlation as in the population of email users ($I_2/I\approx 87\%$), while people's daily and weekly rhythms explain very little of the correlation [$I_C/I\approx 5\%$; Fig. \ref{Ising}(e)].  Interestingly, this difference in $I/I_C$ between email activity and private messages [Fig. \ref{Ising}(c)] reflects the commonly-held intuition that email activity is moderately tied to people's work and leisure schedules, while private messages are not.

We are ultimately interested in understanding the role of pairwise correlations in driving large-scale surges of activity in the entire 100-person population.  With this goal in mind, we calculate the fraction $I_2/I$ in groups of email users increasing in size from $N=2$ through $10$. For small groups and relatively weak correlations, as the group size increases, we expect the multi-information $I$ to increase in proportion to the entropy difference $I_2$ \cite{Schneidman-01}. Indeed, we find that the fraction $I_2/I$ remains nearly constant as the groups grow in size ($I_2/I\propto N^{-0.075\pm 0.005}$). Extrapolating to the entire 100-person population, we find with 95\% confidence that pairwise correlations account for 72-78\% of the total multi-information in the data [Fig. \ref{Ising}(d)]. This fraction is especially large when considering the exponential number of possible higher-order correlations ($\sim 2^N$) for populations of increasing size $N$. We conclude that large-scale patterns of behavior, across several distinct modes of human activity, can be robustly understood as emerging from an underlying network of pairwise correlations.

\section{Modeling an entire population}

\begin{figure*}
\includegraphics[width = .8\textwidth]{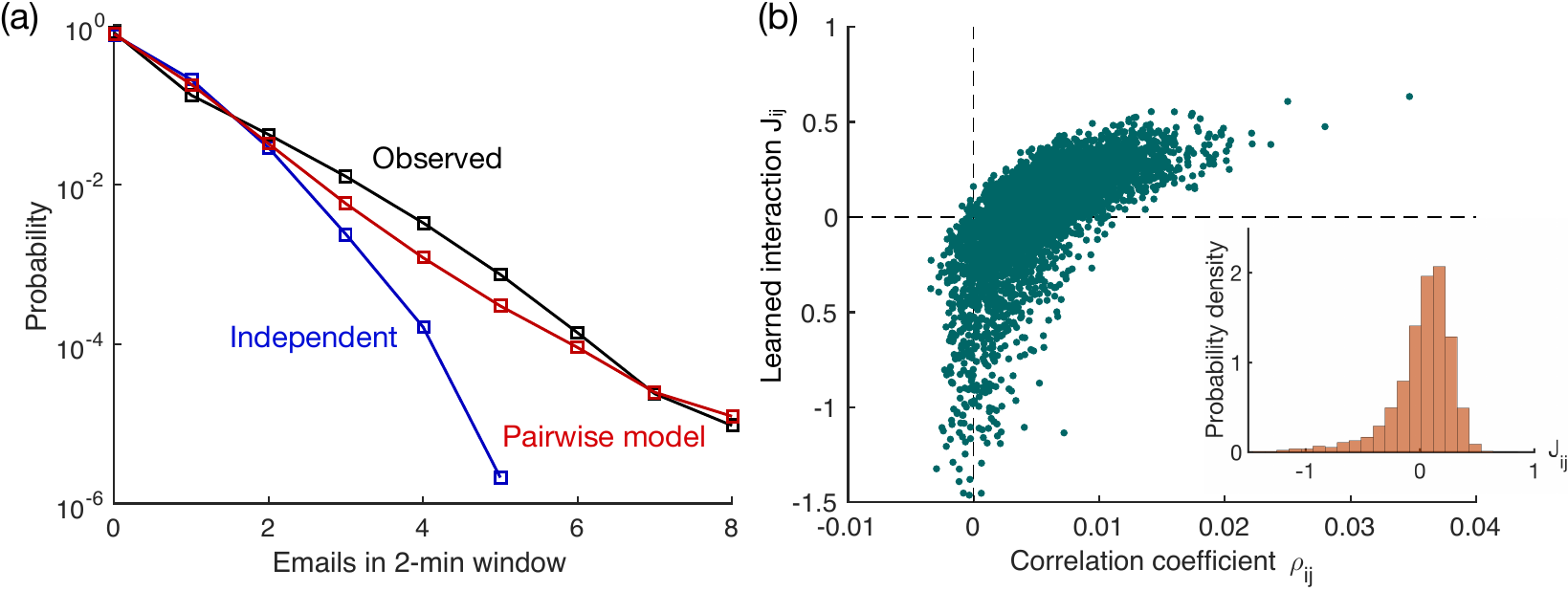}
\caption{\label{Large} Surges of collective activity are captured by pairwise correlations. (a) Distribution of the observed number of emails in a given two-minute window (black), the prediction of the independent model (blue), and the prediction of the pairwise maximum entropy model (red). (b) Scatter plot illustrating the relationship between the observed pairwise correlations in the data $\rho_{ij}$ and the learned Ising interactions $J_{ij}$ for all pairs in the 100-person population. Inset: Histogram of the learned interactions.}
\end{figure*}

Our analysis of relatively small groups indicates that the pairwise maximum entropy model can capture a majority of the correlation structure in groups of up to 100 individuals. This result, in turn, suggests that the heavy-tailed nature of collective human behavior [Fig. \ref{correlations}(d)]---characterized by surges of activity---might emerge organically from pairwise correlations. To test this prediction directly, we must extend the pairwise maximum entropy model to include the entire population of 100 email users. In order to learn the appropriate Ising interactions $J_{ij}$ and external fields $h_i$ for all 100 people, we leverage recent advances in stochastic gradient descent from statistical physics \cite{Ferrenberg-01} and machine learning \cite{Ackley-01}, avoiding the exponential complexity of standard Ising calculations [see Appendix \ref{BM}; Fig. \ref{accuracy}]. Fig. \ref{Large}(a) shows that the pairwise model successfully captures the heavy-tailed nature of human activity, accurately predicting the frequencies of activity surges involving up to seven and eight individuals.

To understand how a network of simple pairwise correlations can generate large-scale spikes in activity, it is useful to study the structure of the Ising parameters in the maximum entropy model [Eq. (\ref{P2})]. We note that each external field $h_i$ either biases individual $i$ toward activity ($h_i > 0$) or toward inactivity ($h_i < 0$). Meanwhile, each Ising interaction $J_{ij}$ either influences individuals $i$ and $j$ to perform actions at the same time ($J_{ij}>0$) or at different times ($J_{ij} < 0$). Here, we draw an important distinction between the learned interactions $J_{ij}$ in the maximum entropy model and the observed pairwise correlations $\rho_{ij}$ in the data: while each pairwise correlation quantifies the frequency with which two individuals perform actions at the same time, each Ising interaction represents a functional influence between two individuals to synchronize their activity, thereby inducing a pairwise correlation. Interestingly, while correlations in the network are weak and almost exclusively positive [Fig. \ref{correlations}(c)], the Ising interactions maintain a large amount of heterogeneity [Fig. \ref{Large}(b), Inset], with almost an equal number of positive and negative interactions. Indeed, the learned pairwise interactions depend highly non-trivially on the corresponding pairwise correlations in the data [Fig. \ref{Large}(b)]. Importantly, the presence of competing positive and negative interactions generates ``frustration,'' as in spin glasses \cite{Mezard-01}, wherein triplets of individuals cannot find a combination of activity and inactivity that simultaneously satisfies all of their interactions. This frustration gives rise to a complex energy landscape of activity patterns with many different local minima, some of which correspond to patterns involving many more active individuals than would be expected under the independent model, thus giving rise to the heavy-tailed behavior in Fig. \ref{Large}(a). Intriguingly, such frustrated interactions have previously been hypothesized to drive a number of social phenomena \cite{Castellano-01}, such as the formation of coalitions \cite{Galam-06}. By calculating the specific Ising parameters that describe each population, and by identifying the presence of competing positive and negative interactions [Fig. \ref{Large}(b), Inset], our work provides rigorous evidence for these long-standing hypotheses.

\begin{figure*}
\includegraphics[width = .85\textwidth]{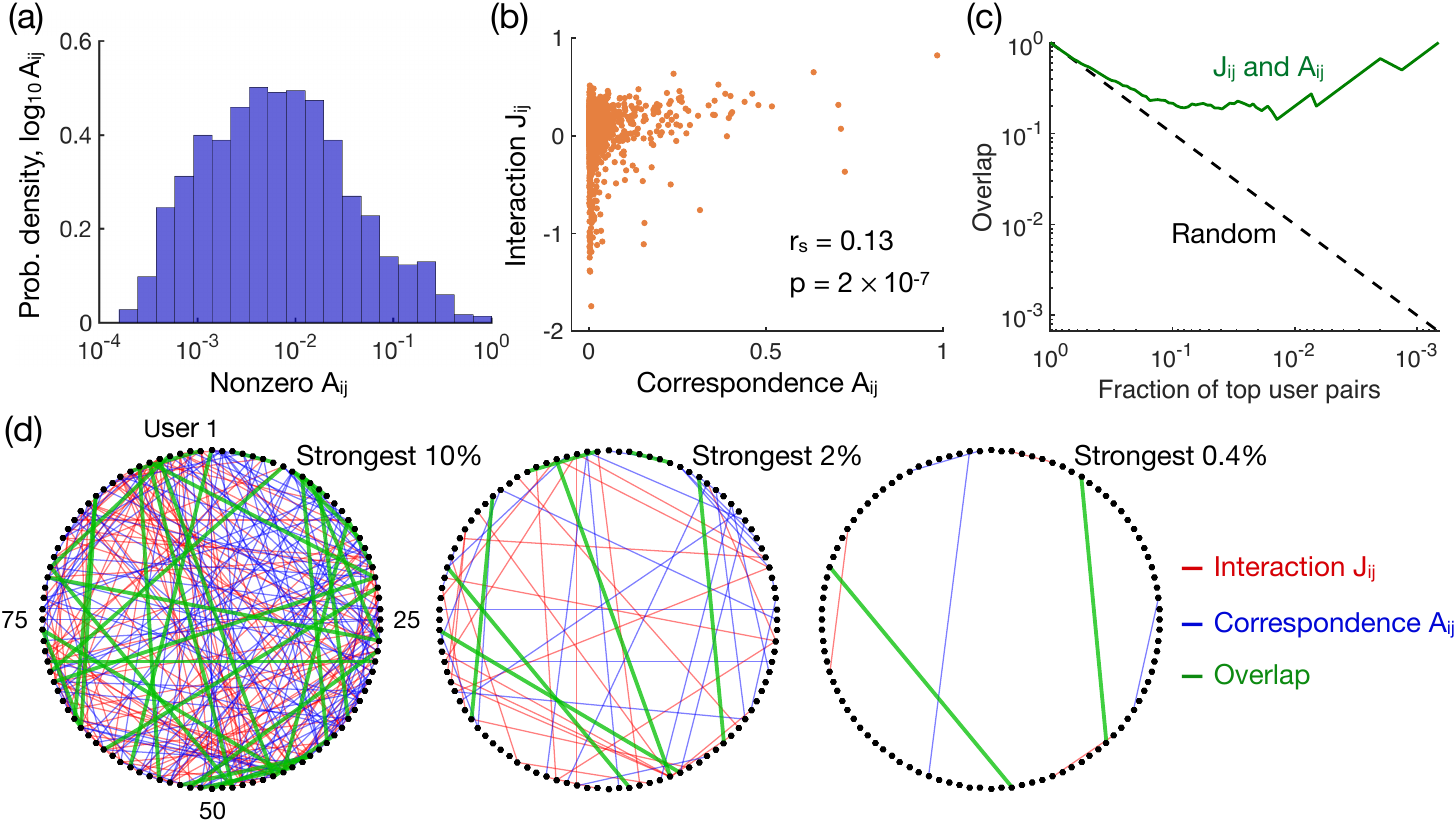}
\caption{\label{Overlap} The learned pairwise interactions uncover pathways of ground truth communication. (a) Histogram of correspondence rates $A_{ij}$ between all pairs of individuals that exchanged at least one email. (b) Scatter plot of the learned Ising interactions versus email correspondence rates for pairs that exchanged at least one email. Importantly, $J_{ij}$ and $A_{ij}$ are significantly correlated with Spearman's correlation coefficient $r_s=0.13$ ($p=2\times 10^{-7}$). (c) Overlap between the strongest interactions $J_{ij}$ and most frequently corresponding pairs $A_{ij}$ as a function of the fraction of pairs being considered. The dashed line indicates the overlap with a random selection of user pairs. (d) Structure of the strongest pairwise interactions (red), highest correspondence rates (blue), and overlap between the two (green) for all 100 individuals. The three networks represent the strongest 10\% (left), 2\% (middle), and 0.4\% (right) of user pairs.}
\end{figure*}

\section{The role of inter-human communication}

Thus far, we have focused on understanding correlations in the timing of actions, without knowledge of who each person is interacting with in the population. Fundamentally, the Ising interactions $J_{ij}$ are merely learned parameters that ensure consistency with the observed pairwise correlations in the network. However, it is tempting to imbue them with physical significance, interpreting these functional interactions as comprising a network of real-world influences between individuals. For previous applications of maximum entropy models in neuroscience \cite{Schneidman-01, Ganmor-01} and biology \cite{Bialek-01,Weigt-01,Lezon-01}, because comparisons with ground truth interactions are often infeasible, any physical meaning attributed to the learned interactions $J_{ij}$ has remained, at its core, an analogy. By contrast, in the context of email activity, we automatically know a subset of the ground truth interactions---namely, the network of email communication between individuals. Although it is appealing to suspect that the learned Ising interactions are closely related to the structure of email correspondence in the data, we emphasize that this need not be the case. There is an array of circumstances that could influence the activity of two individuals to become correlated, from common functional roles in the network to shared communication with an external third party. Furthermore, even if correlations do arise from direct communication, this communication could take on many forms that do not appear in the dataset, including face-to-face contact, texts, calls, or other online avenues.

Keeping in mind these reasons for guarded optimism, here we compare the learned interactions $J_{ij}$ from our maximum entropy model with the network of email traffic between individuals. Letting $n_{i\rightarrow j}$ denote the number of emails sent from person $i$ to person $j$, and letting $n_i = \sum_j n_{i\rightarrow j}$ denote the total number of emails sent by person $i$, we define the correspondence rate between two people $i$ and $j$ to be $A_{ij} = (n_{i\rightarrow j} + n_{j\rightarrow i})/(n_i + n_j)$. In words, $A_{ij}$ represents the fraction of the $n_i+n_j$ emails sent by person $i$ and person $j$ that were addressed to each other. We find that most correspondence only accounts for around 1\% of a pair's total email communication, while a small number of pairs communicate almost exclusively with one another [Fig. \ref{Overlap}(a)]. Considering all pairs of people that exchanged at least one email ($A_{ij} > 0$), we find that the learned Ising interactions $J_{ij}$ are significantly correlated with the correspondence rates $A_{ij}$ in the data [Spearman's correlation coefficient $r_s = 0.13$, $p=2\times 10^{-7}$; Fig. \ref{Overlap}(b)]. This relationship between the learned Ising interactions and the ground truth communication in the population is particularly interesting after reflecting on the myriad ways in which these two networks could have remained unrelated, as described above.

To fully appreciate the strength of the relationship between $J_{ij}$ and $A_{ij}$, we focus on the fraction $f$ of the strongest pairwise interactions and correspondence rates in the population. These two thresholded networks overlap significantly [Fig. \ref{Overlap}(c)], with the strongest 1\% of Ising interactions exhibiting a 20\% overlap with the top 1\% of frequently communicating pairs---20 times higher than if $J_{ij}$ and $A_{ij}$ were independent. This overlap becomes even more pronounced as we increase the threshold [Fig. \ref{Overlap}(d)], such that the single strongest maximum entropy interaction in the entire population corresponds precisely to the pair of individuals that communicate most frequently. This relationship between $J_{ij}$ and $A_{ij}$ provides a compelling mechanistic interpretation for the Ising interactions in our maximum entropy model; namely, frequent communication between a pair of individuals (quantified by $A_{ij}$) acts as an influence to synchronize their activity (quantified by $J_{ij}$). As demonstrated in previous sections, the resulting pairwise correlations, in turn, can generate the types of large-scale correlations and surges in human activity that are ubiquitous in the modern world \cite{Sornette-01,Deschatres-01,Candia-01,Crane-01,Peng-01,Bagrow-01}.

\section{Conclusions and future directions}

Despite the widespread investigation of fine-scale correlations as the building blocks of large-scale behavior in complex systems throughout physics \cite{Jaynes-01,Cover-01}, neuroscience \cite{Schneidman-01, Ganmor-01}, and biology \cite{Bialek-01,Weigt-01,Lezon-01}, a similar quantitative approach to human dynamics has been notably lacking. Here, we provide an important step toward the ultimate goal of understanding the role of fine-scale correlations in generating large-scale patterns of human activity. Studying four datasets that reflect the diversity of human activity, we first show that all populations exhibit surges of collective activity, a phenomenon that has become the subject of intense research focus \cite{Sornette-01,Deschatres-01,Candia-01,Crane-01,Peng-01,Bagrow-01}. Importantly, these surges in activity cannot be accounted for by commonly-used models that assume independence in human behavior \cite{Haight-01,Rasch-01,Karagiannis-01,Gerlough-01}. To understand where surges in activity come from, we consider the possibility that large-scale patterns arise naturally from combinations of simple pairwise correlations between individuals. To formalize this hypothesis, we utilize the principle of maximum entropy from information theory, deriving a pairwise maximum entropy model of human activity that is formally equivalent to an Ising model. Interestingly, this maximum entropy model accounts for 72-78\% of the total correlation in a 100-person population of email users [Fig. \ref{Ising}(e)] and accurately predicts the heavy-tailed distribution of activity surges [Fig. \ref{Large}(a)]. Additionally, we demonstrate that the Ising interactions in our model closely resemble the network of inter-human communication within the population. This close relationship between functional interactions and ground truth communication suggests an intuitive mechanism driving pairwise correlations.

Just as emergent phenomena have garnered significant attention in the natural sciences \cite{Jaynes-01,Cover-01,Schneidman-01,Marre-01,Ganmor-01,Bialek-01,Weigt-01,Lezon-01}, we anticipate that similar approaches will prove fruitful in the development of accurate models of large social systems. Importantly, while a majority of existing research has focused on the impacts that external influences have on human populations \cite{Peng-01,Bagrow-01}, these explanations are fundamentally limited by their reliance on context-specific mechanisms \cite{Crane-01,Candia-01}. By contrast, interactions between humans are present in almost every context, and, as we have demonstrated, these interactions can build upon one another to have a large-scale impact on the behavior of an entire population. In this way, thinking carefully about the role of fine-scale correlations in activity can have quite general implications for resource allocation in communication \cite{Candia-01} and transportation \cite{Peng-01} networks, understanding social organization \cite{Onnela-01}, and preventing viral epidemics \cite{Pastor-01}.

To conclude, we point out a number of limitations of our analysis that highlight important directions for future work. First, we remark that, given the diversity of experiences that shape human actions, it would be na\"{i}ve to conclude that all collective behaviors only emerge from internal correlations. To the contrary, it has been well established that external influences play an important role in predicting a number of collective human behaviors \cite{Sornette-01,Deschatres-01,Candia-01,Crane-01,Peng-01,Bagrow-01}. Therefore, future work should investigate the interplay between external influences and internal interactions in human populations. Such an investigation would likely benefit from advances in control theory and influence maximization \cite{Kempe-01,Morone-01}, which have recently been used to predict the propagation of external influences in Ising networks \cite{Lynn-01, Lynn-02, Lynn-03}. Second, we note that our investigation has focused primarily on pairwise correlations. While these simplest correlations represent a logical first step, our results do not rule out the possibility that higher-order correlations could also have an important impact on large-scale behavior. Practically speaking, the primary difficulty in studying such higher-order correlations lies in determining which to include in a maximum entropy model, as there exist ${N}\choose{k}$ different choices for each $k^{\text{th}}$-order correlation (a number that grows nearly exponentially with $k$). Fortunately, to handle this explosion of parameters, recent advances in neuroscience have produced tractable techniques for generating sparse higher-order maximum entropy models \cite{Ganmor-01}. Such higher-order models represent systematic generalizations of the methods presented here, and could prove vital for understanding the large-scale impacts of triplet and quadruplet correlations [Fig. \ref{Influence}(b)], which are thought to encode important organizational features in human populations \cite{Eckmann-01} (see Appendix \ref{Discussion} for an extended discussion).

\begin{acknowledgments}
D.S.B. and C.W.L. acknowledge support from the John D. and Catherine T. MacArthur Foundation, the Alfred P. Sloan Foundation, the ISI Foundation, the Paul Allen Foundation, the Army Research Laboratory (W911NF-10-2-0022), the Army Research Office (Bassett-W911NF-14-1-0679, Grafton-W911NF-16-1-0474, DCIST- W911NF-17-2-0181), the Office of Naval Research, the National Institute of Mental Health (2-R01-DC-009209-11, R01-MH112847, R01-MH107235, R21-M MH-106799), the National Institute of Child Health and Human Development (1R01HD086888-01), National Institute of Neurological Disorders and Stroke (R01 NS099348), and the National Science Foundation (BCS-1441502, BCS-1430087, NSF PHY-1554488 and BCS-1631550). L.P. is supported by an NSF Graduate Research Fellowship. D.D.L. acknowledges support from the NSF, NIH, DOT, ARL, AFOSR, ONR and DARPA. The content is solely the responsibility of the authors and does not necessarily represent the official views of any of the funding agencies. 
\end{acknowledgments}

\appendix

\section{Data preprocessing}

\label{Data}

\begin{figure}[h]
\includegraphics[width = .35\textwidth]{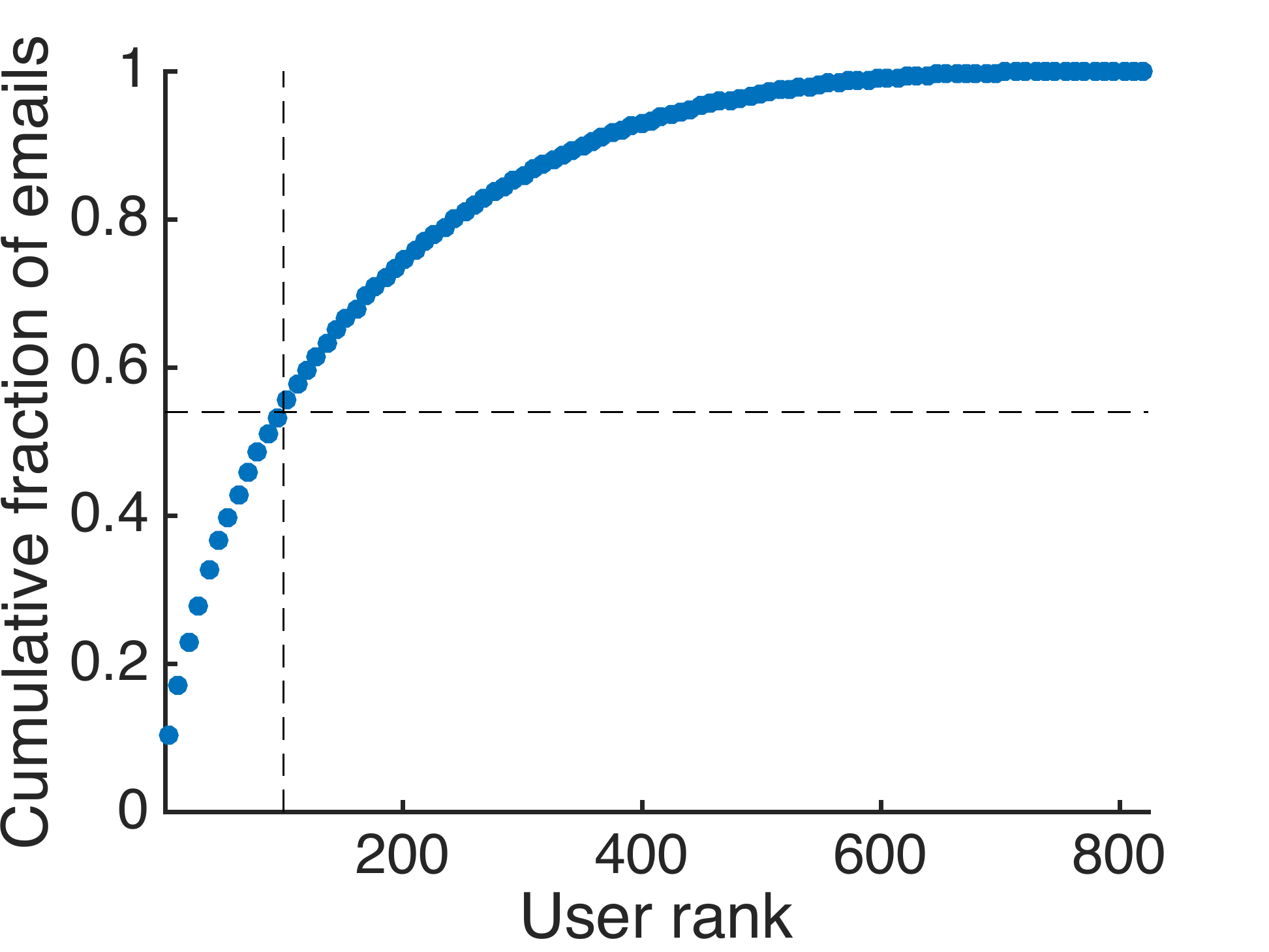}
\caption{\label{ParamChoices} Cumulative distribution of emails versus the activity rank of the users. The 100 most active individuals account for 56\% of the emails in the network (dashed lines).}
\end{figure}

Here, we discuss the details of how the email data is processed, noting that the other datasets follow in an analogous fashion. In total, the dataset contains the email correspondence between 986 members of a European research institution over 526 days \cite{Paranjape-01}. We focus on the 100 most active individuals, roughly corresponding to the members of the population that sent on average at least one email per day [Fig. \ref{ParamChoices}]. To quantify correlations between different individuals, we must discretize the data into time bins of width $\Delta t$.
\begin{figure*}
\includegraphics[width = \textwidth]{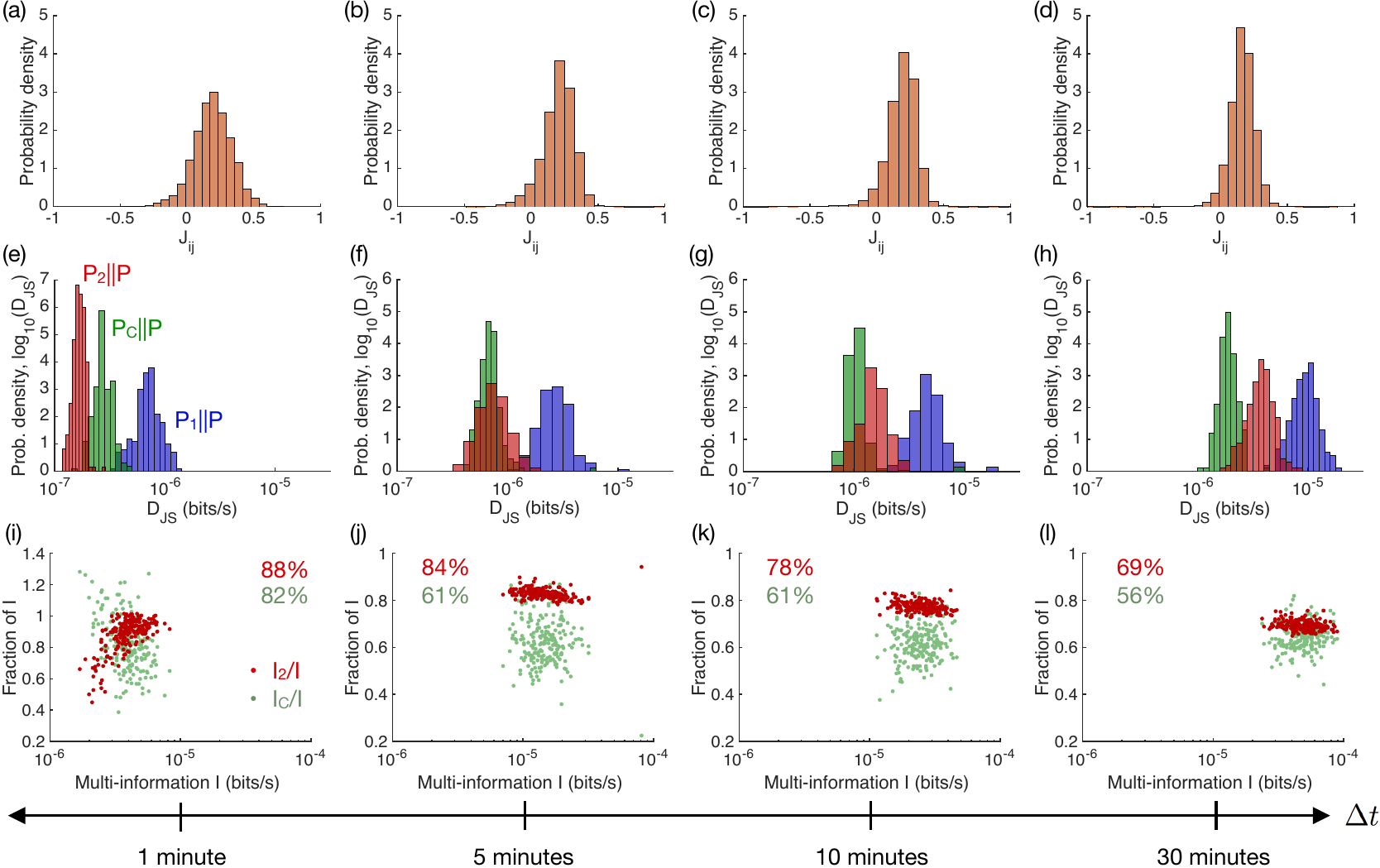}
\caption{\label{ChangeDt} Dependence of the pairwise maximum entropy model on the bin width. (a-d) Distributions of pairwise couplings for 200 different 10-person groups selected from the 100 most active individuals in the email dataset. From left to right, the data is discretized into bins of width $\Delta t = $ 1, 5, 10, and 30 minutes. (e-h) Jensen-Shannon divergences between the observed distribution over activity patterns $P$ and the independent $P_1$ (blue), maximum entropy $P_2$ (red), and conditionally independent $P_C$ (green) models. The distributions are taken over the 200 groups from panels (a-d). (i-l) Fraction of the network correlation captured by the maximum entropy (red) and conditionally independent (green) models, plotted against the full network correlation, quantified by the multi-information $I$. The average percentage of the multi-information captured by each model is displayed in the upper corner. Each dot represents a different group of 10 people, and $I$ is divided by $\Delta t$ to remove dependence on the window size.}
\end{figure*} 
To choose a suitable bin width, we notice that 90\% of consecutive emails from the same individual are sent with at least two minutes in between [Fig. \ref{correlations}(a)], defining a natural time scale that we use as our $\Delta t$.Discretizing the 526-day dataset into 2-minute bins, we produce a set of $\sim 3.8\times 10^5$ binary patterns $\{\bm{\sigma}\}$ that define the behavior of our population.

\section{Robustness of the pairwise model}

\label{Robustness}

In Appendix \ref{Data}, we provided first-principles justifications for focusing on the 100 most active individuals in the email dataset and for discretizing the data into bins of width $\Delta t = 2$ minutes. Here, we verify that the success of the pairwise maximum entropy model is robust to reasonable variations in these choices.

\begin{figure*}
\includegraphics[width = .85\textwidth]{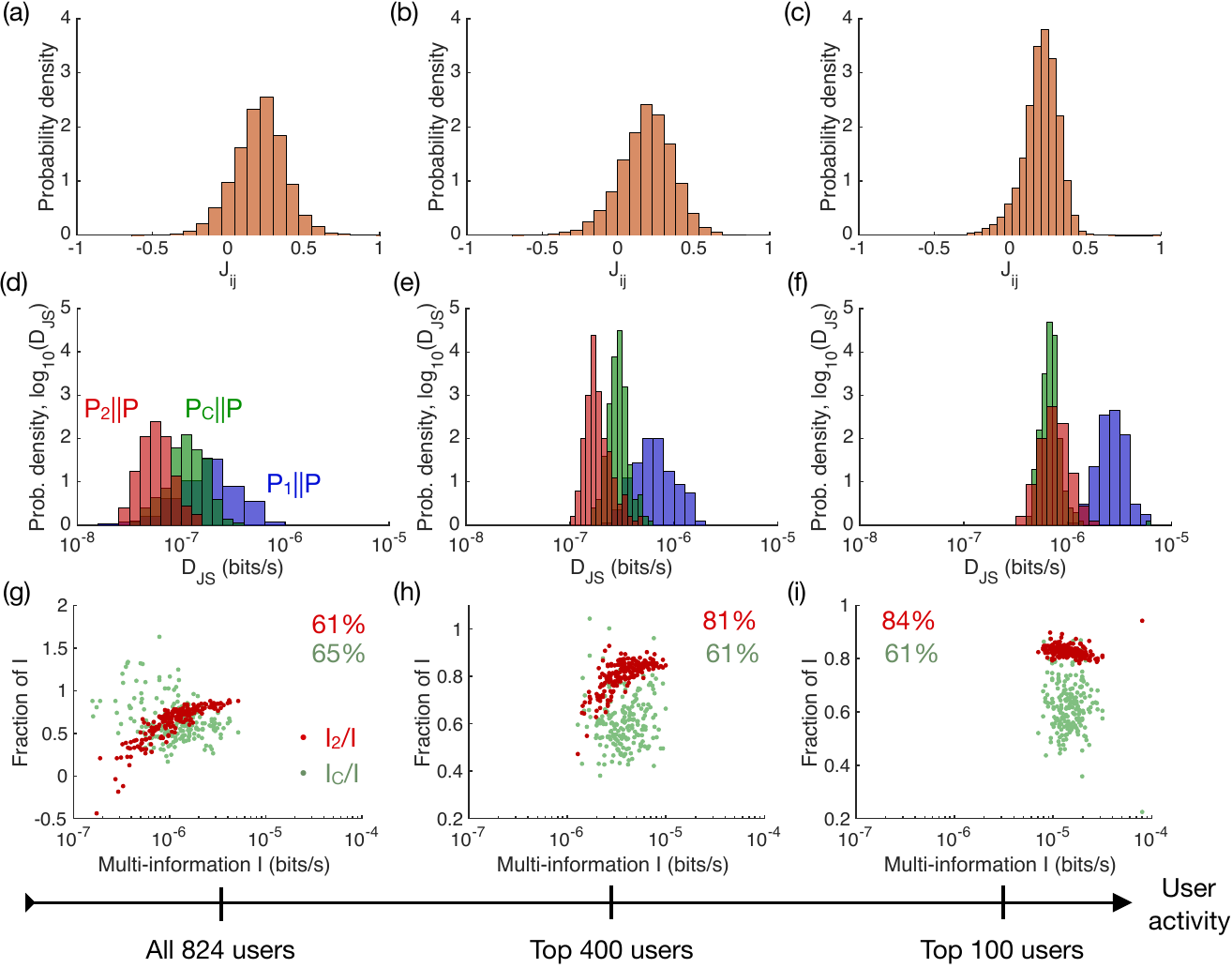}
\caption{\label{ChangeUsers} Dependence of the pairwise model on the set of individuals chosen for analysis in the email dataset. (a-c) Distributions of pairwise interactions for 200 different groups of 10 individuals, where the data is discretized with bin width $\Delta t = 5$ minutes. From left to right, the 200 groups are chosen from among all 824 people that sent at least one email, the 400 most active individuals, and the 100 most active individuals, respectively. (d-f) Jensen-Shannon divergences between the observed distribution over activity patterns $P$ and the independent $P_1$ (blue), maximum entropy $P_2$ (red), and conditionally independent $P_C$ (green) models. The distributions are taken over the 200 groups of users. (g-i) Fraction of the network correlation captured by the pairwise maximum entropy (red) and conditionally independent (green) models, plotted against the full network correlation, quantified by the multi-information $I$. The average percentage of the multi-information captured by each model is displayed in the upper corner. The multi-information is divided by $\Delta t$ to remove dependence on the window size.}
\end{figure*}

\begin{figure*}
\includegraphics[width = .8\textwidth]{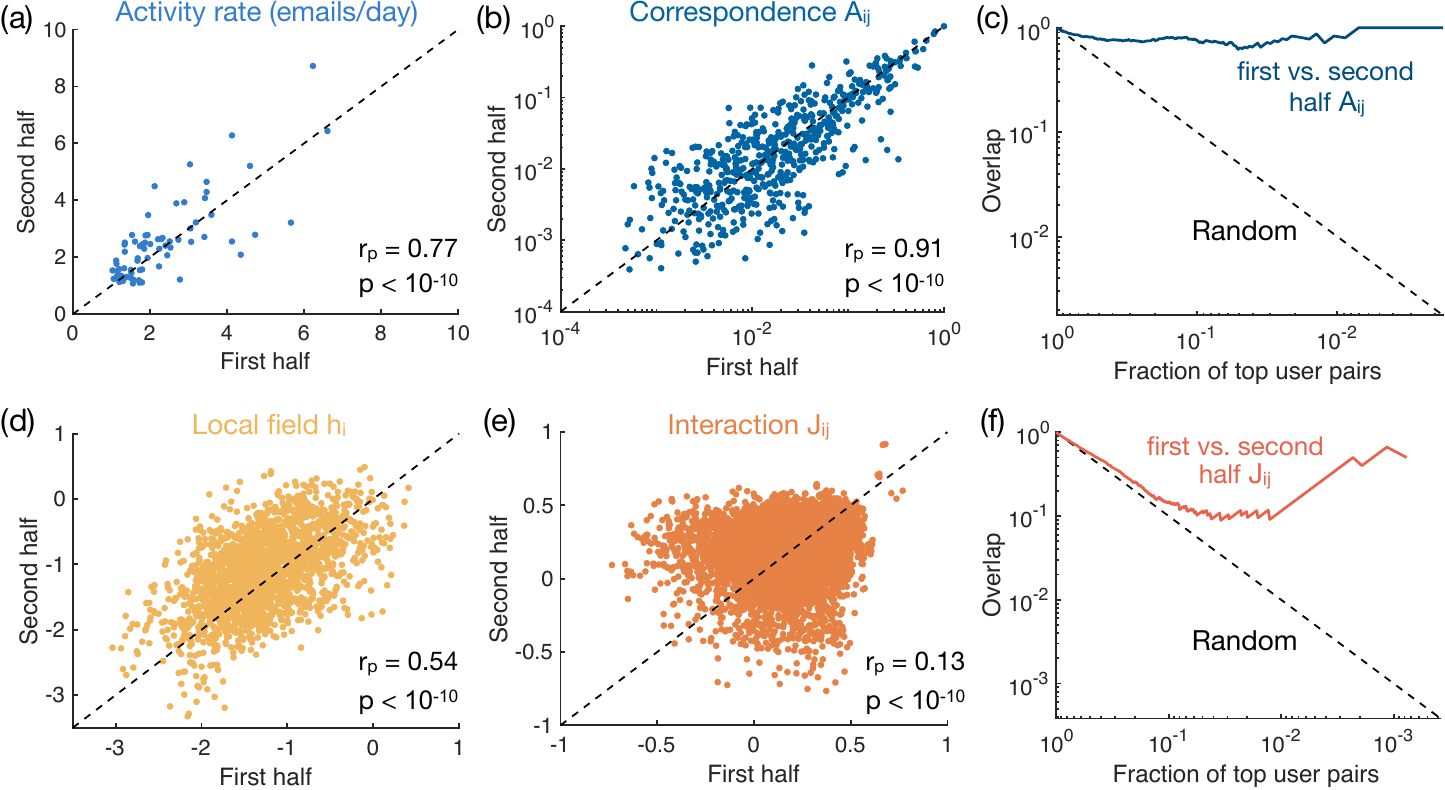}
\caption{\label{TimeComp} Consistency of the pairwise maximum entropy model over time. (a) Comparison of email user activity rates in the first half versus the second half of the dataset; the dashed line indicates equality. (b) Correspondence rates $A_{ij}$ between pairs of users are strongly correlated across the two halves of the dataset. (c) Overlap between the most frequently corresponding pairs of users in the first half and those in the second half as a function of the fraction of pairs being considered. The dashed line indicates the overlap with a random selection of user pairs. (d) For 200 random groups of 10 individuals, we compare the local fields $h_i$ of pairwise maximum entropy models fit to either the first or second half of the email data. (e) For the same 200 random groups, we compare the Ising interactions $J_{ij}$ of the pairwise models fit to the two halves of the dataset. (f) For each half of the dataset, we average the interactions $J_{ij}$ over all 200 groups and plot the overlap between average interaction networks as a function of the fraction of user pairs being considered. As in panel (c), the dashed line indicates the overlap with a random selection of pairs.}
\end{figure*}

\subsection{Dependence on the bin width}

\label{Dt}

We investigate the dependence of the pairwise maximum entropy model on the bin width $\Delta t$ used to discretize the email activity. Throughout, we focus on the 100 most active individuals, and we consider bin widths of $\Delta t = $ 1, 5, 10, and 30 minutes. For each value of $\Delta t$, we randomly select 200 different groups of 10 individuals and fit a pairwise maximum entropy model to describe each group. As $\Delta t$ increases, we witness more windows involving multiple active individuals, thereby strengthening the correlations that we observe in the discretized data. In turn, these stronger correlations give rise to Ising interactions $J_{ij}$ that are more positive and sharply peaked [Fig. \ref{ChangeDt}(a-d)]. In Fig. \ref{ChangeDt}(e-h), we show that the true distribution of activity is approximately five times closer to the maximum entropy model $P_2$ than to the independent model $P_1$ across all values of $\Delta t$ considered, demonstrating the consistency of the pairwise model in predicting human behavior. On the other hand, the performance of the conditionally independent model $P_C$ increases significantly as $\Delta t$ increases, even outperforming the pairwise model for $\Delta t \ge 10$ minutes. We note, however, that for such large bin widths, people often send multiple emails within the same window, and treating the data as binary may not be justified. In Fig. \ref{ChangeDt}(i-l), we see that the pairwise model captures nearly all of the multi-information in the 10-person groups across all choices for $\Delta t$. By contrast, the conditionally independent model consistently captures a smaller fraction of the multi-information in the data. Furthermore, for $\Delta t = 1$ minute, the conditionally independent model has lower entropy than the data itself (i.e., $I_C/ I > 1$) for 30 of the 200 groups, which is a clear indication that the model is overfitting the data.

\subsection{Dependence on the individuals being analyzed}

\label{Users}

We investigate the dependence of the maximum entropy model on the set of individuals chosen for analysis. In particular, we consider 200 different 10-person groups selected from among the 100 most active email users, the 400 most active users, and all 824 users that sent at least one email. Throughout this section, the bin width is fixed at $\Delta t = 5$ minutes. As we focus on more active individuals, the observed correlations become stronger, which is reflected in the fact that the distribution of learned interactions $J_{ij}$ among the top 100 individuals is more sharply peaked and positive than the pairwise interactions between the top 400 and all 824 individuals [Fig. \ref{ChangeUsers}(a-c)]. In Fig. \ref{ChangeUsers}(d-f), we again find that the pairwise model is approximately five times closer to the true distribution than the independent model across all three subpopulations. By contrast, the conditionally independent model performs nearly as well as the pairwise model among the 100 most active individuals, but provides only marginal improvements over the independent model for all 824 individuals. The failure of the conditionally independent model in describing the entire 824-person population is not surprising given that most individuals sent less than one email every five days, leaving daily and weekly rhythms with little to no predictive power.

We now consider the fraction of the multi-information captured by each model. For all 824 individuals, Fig. \ref{ChangeUsers}(g) shows that the conditionally independent model captures a slightly larger fraction of the multi-information than the maximum entropy model; however, $P_C$ erroneously includes more correlation than the data itself (i.e., $I_C/I > 1$) for 20 of the 200 groups of 10 people, indicating that the model is overfitting the data. For both the top 100 and 400 most active individuals, the maximum entropy model captures a significantly larger fraction of the network correlation than the conditionally independent model [Fig. \ref{ChangeUsers}(h-i)].

We conclude that the predictions of the pairwise maximum entropy model are robust to variations in both the bin width $\Delta t$ as well as the set of individuals chosen for analysis.

\begin{figure*}
\includegraphics[width = .85\textwidth]{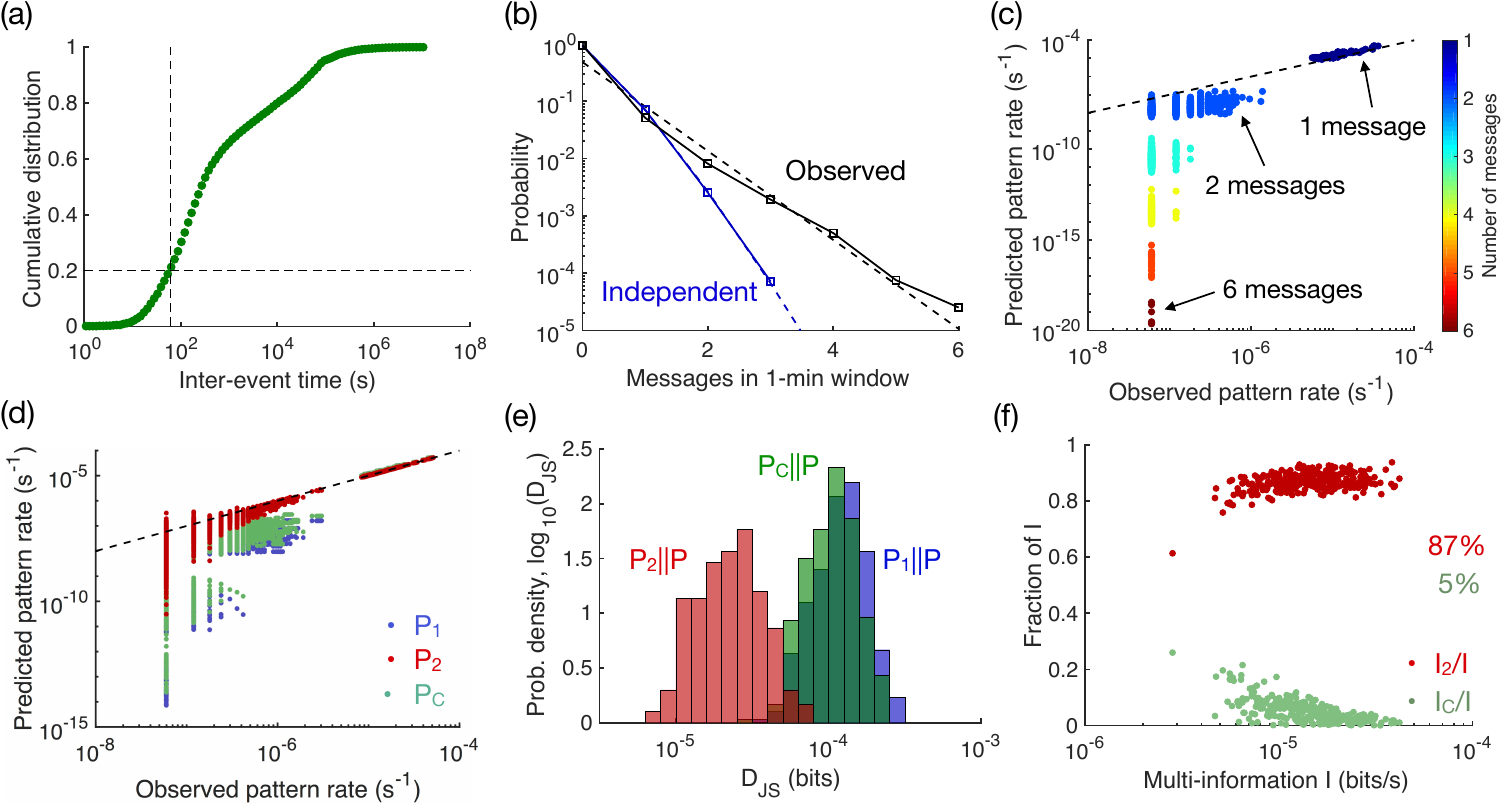}
\caption{\label{Irvine_fig} Performance of the pairwise maximum entropy model in a dataset of private messages. (a) Cumulative distribution of inter-event times for the 66 most active individuals. Approximately 80\% of consecutive messages from the same person are sent with at least one minute in between (dashed lines). (b) Distribution of the messages sent in a given one-minute window in the dataset (black) and after shuffling individuals' activities to eliminate correlations (blue); dashed lines indicate an exponential fit to the observed data (black) and a Poisson fit to the shuffled data (blue). (c) The rate of each observed activity pattern, plotted against the approximate rate under the independent model $P_1$; the dashed line indicates equality. (d) We plot the rate of each observed activity pattern across 300 randomly selected groups of 10 individuals against the approximate rates under the independent model $P_1$ (blue), the pairwise maximum entropy model $P_2$ (red), and the conditionally independent model $P_C$ (green); the dashed line indicates equality. (e) Jensen-Shannon divergences between the true distribution $P$ and the independent $P_1$ (blue), maximum entropy $P_2$ (red), and conditionally independent $P_C$ (green) models; the histograms reflect estimates from the 300 10-person groups. (f) Fraction of the network correlation (i.e., multi-information $I$) captured by the pairwise (red) and conditionally independent (green) models, plotted against the full multi-information. We note that $I$ is divided by $\Delta t$ to remove the dependence on window size.}
\end{figure*}

\subsection{Consistency of the pairwise model over time}

\label{Consistency}

By employing the pairwise maximum entropy model in Eq. (\ref{P2}), we implicitly assume that the population activity can be modeled as a stationary distribution; that is, that the local fields $h_i$ and interactions $J_{ij}$ do not change over time. Here, we test this assumption explicitly while noting that the development of time-evolving maximum entropy models is an important direction for future work (see Appendix \ref{Dynamic} for an extended discussion). Specifically, we wish to determine if the Ising parameters describing one portion of the email activity resemble those describing another portion of the activity. To do so, we divide the dataset into two halves corresponding roughly to the first and last 263 days of email activity. Fig. \ref{TimeComp}(a-c) shows that the statistics describing the population activity remain remarkably consistent over time, with both the user activity rates and pair correspondence rates $A_{ij}$ being strongly correlated between the two halves of data (Pearson's correlations $r_p = 0.77$ for the activity rates and $r_p = 0.91$ for the correspondence rates).

To study the consistency of the maximum entropy model, we randomly select 200 different 10-person groups from among the 74 users that sent at least one email in both halves of the dataset, and we then learn pairwise models describing each group for each half of data. Fig. \ref{TimeComp}(d-e) shows that the local fields $h_i$ and interactions $J_{ij}$ modeling the population activity are significantly correlated over time (Pearson's correlations $r_p = 0.54$ for the local fields and $r_p = 0.13$ for the interactions). The consistency of the Ising interactions $J_{ij}$ between the two halves of data becomes even more apparent when we focus on the strongest interactions in the population [Fig. \ref{TimeComp}(f)]. Together, these results indicate that the patterns of population activity remain relatively consistent over time, justifying our application of the stationary maximum entropy model as a first step toward more complex dynamical models.

\section{Other modes of human activity}

In the main text, our analysis focused primarily on a dataset of email activity. Here, we independently verify the ability of the pairwise maximum entropy model to quantitatively describe collective human behavior in three other datasets representing a diverse range of human activities.

\subsection{Private messages}

\label{Irvine}

We first consider a dataset of $\sim 6\times 10^5$ private messages sent between 1899 students at U.C. Irvine over the span of 193 days \cite{Panzarasa-01}. As in the context of email activity, we focus on the individuals that sent on average at least one message per day, corresponding to the 66 most active students in the population. To choose an appropriate bin width, we consider the distribution of time gaps between consecutive messages from the same student [Fig. \ref{Irvine_fig}(a)]. Comparing against the equivalent distribution in the email dataset [Fig. \ref{ParamChoices}(b)], we notice that many more private messages than emails are sent with short gaps ($\lesssim 1$ minute) in between. This bursty behavior indicates that the private messages serve as a more conversational communication medium than emails, a fact that will later help in understanding the impact of daily and weekly rhythms. Due to the bursty nature of private messages, we reduce our bin width to $\Delta t = 1$ minute, yielding a dataset of $\sim 2.8\times 10^5$ binary activity patterns.

As in the network of email correspondence, the independent model $P_1$ fails to explain the collective behavior in the private message population \cite{Rasch-01,Karagiannis-01,Haight-01,Gerlough-01}; while the independent model predicts a super-exponential drop off in the number of active individuals in a given window, we find that the distribution of private messages is actually heavy-tailed, fitting closely to an exponential distribution [Fig. \ref{Irvine_fig}(b)]. Additionally, in Fig. \ref{Irvine_fig}(c) we see that the independent model dramatically under-predicts patterns involving two or more active individuals. To improve upon the independent model, we again consider two competing hypotheses: (i) that large-scale patterns emerge from an aggregation of simple pairwise correlations (represented by the pairwise maximum entropy model $P_2$), and (ii) that large-scale patterns are driven by similarities in people's weekly routines (represented by the conditionally independent model $P_C$). Randomly selecting 300 groups of 10 people, Fig. \ref{Irvine_fig}(d) shows that the pattern rates predicted by the pairwise maximum entropy model are tightly correlated with the observed pattern rates, avoiding the inaccuracies of the independent and conditionally independent models.

Additionally, calculating the Jensen-Shannon divergences $D_{JS}(Q||P)$ from each model $Q$ to the observed data $P$, we find that one would typically need over five times more samples to distinguish the pairwise model than the independent model [Fig. \ref{Irvine_fig}(e)], reflecting roughly the same performance as in the network of email correspondence. Interestingly, in contrast to email activity, the conditionally independent model provides nearly no improvement over the independent model in the dataset of private messages. Additionally, Fig. \ref{Irvine_fig}(f) shows that the pairwise maximum entropy model captures $I_2/I \approx 87\%$ of the correlation in the data, nearly identical to its performance on the network of email correspondence, while the conditionally independent model accounts for a strikingly small fraction of the correlation structure ($I_C/I \approx 5\%$). This difference in the performance of $P_C$ between the private message and email datasets suggests that the conversational nature of private messages makes them less likely than email traffic to depend on people's routines. By contrast, the maximum entropy model accurately describes the activity in both populations, further validating the conclusion that patterns of collective behavior can be understood as emerging from simple pairwise correlations.

\begin{figure}
\includegraphics[width = .49\textwidth]{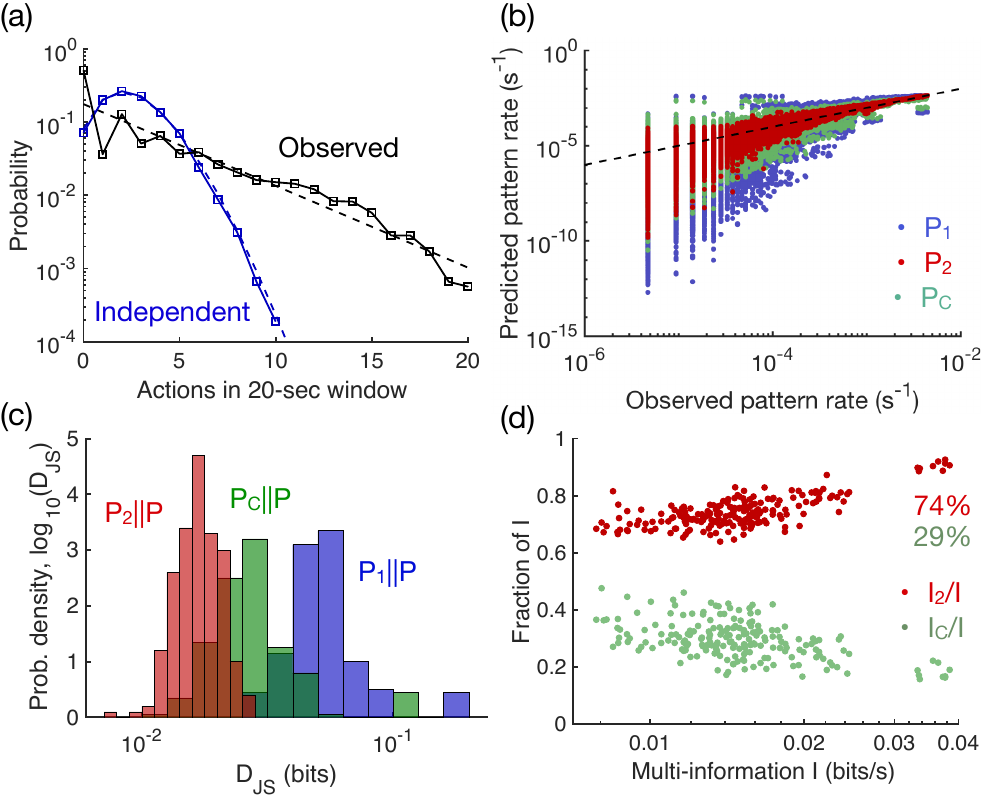}
\caption{\label{Conference_fig} Performance of the pairwise model in a dataset of face-to-face contacts between individuals. (a) Distribution of the number of contacts in a given 20-second window observed in the dataset (black) and after shuffling individuals' activities to eliminate correlations (blue); dashed lines indicate an exponential fit to the observed data (black) and a Poisson fit to the shuffled data (blue). (b) The rate of each observed activity pattern across 200 randomly selected groups of 10 individuals is plotted against the approximate rates under the independent model $P_1$ (blue), the pairwise maximum entropy model $P_2$ (red), and the conditionally independent model $P_C$ (green); the dashed line indicates equality. (c) Jensen-Shannon divergences between the true distribution $P$ and the independent $P_1$ (blue), maximum entropy $P_2$ (red), and conditionally independent $P_C$ (green) models; the histograms reflect estimates from the 200 10-person groups. (d) Fraction of the network correlation (i.e., multi-information $I$) captured by the pairwise (red) and conditionally independent (green) models, plotted against the full multi-information; $I$ is divided by $\Delta t = $ 20 seconds to remove the dependence on window size.}
\end{figure}

\subsection{Physical contacts}

\label{Conference}

Thus far, we have only studied human actions mediated by online communication. Here, we instead consider a dataset of face-to-face interactions between 50 attendees at the ACM Hypertext 2009 conference, which spanned three days \cite{Isella-01}. Discretizing the population activity into bins of width $\Delta t = $ 20 seconds, we arrive at a set of $\sim 10^4$ binary activity vectors. As in both the networks of email and private message correspondence, we observe that the number of human contacts within a given 20-second window roughly obeys an exponential distribution, while the independent model instead predicts a Poisson distribution that severely under-predicts the likelihood of surges in human activity [Fig. \ref{Conference_fig}(a)]. To study the pairwise maximum entropy model, we generate 200 random groups of 10 individuals. Fig. \ref{Conference_fig}(b) shows that the rates of activity patterns predicted by the pairwise model are tightly correlated with the rates at which they were observed at the conference, providing consistently more accurate predictions than both the independent  and conditionally independent models.

Quantitatively, one would require three to four times as many samples to distinguish the independent model from the observed data than the maximum entropy model, and the maximum entropy model achieves a lower Jensen-Shannon divergence from the observed data than the conditionally independent model across all 200 groups of attendees [Fig. \ref{Conference_fig}(c)]. Additionally, Fig. \ref{Conference_fig}(d) shows that the pairwise model captures $I_2/I \approx 74\%$ of the correlation in the face-to-face contacts. While this is slightly lower than that observed for emails and private messages, we remark that the conditionally independent model only accounts for $I_C/I \approx 29\%$ of the correlation in the data. Interestingly, despite physical interactions representing a quite different mode of human activity from online communication, we still find that patterns of population behavior are well-described as arising from pairwise correlations.

\subsection{Music streams}

\label{LastFM}

\begin{figure}
\includegraphics[width = .49\textwidth]{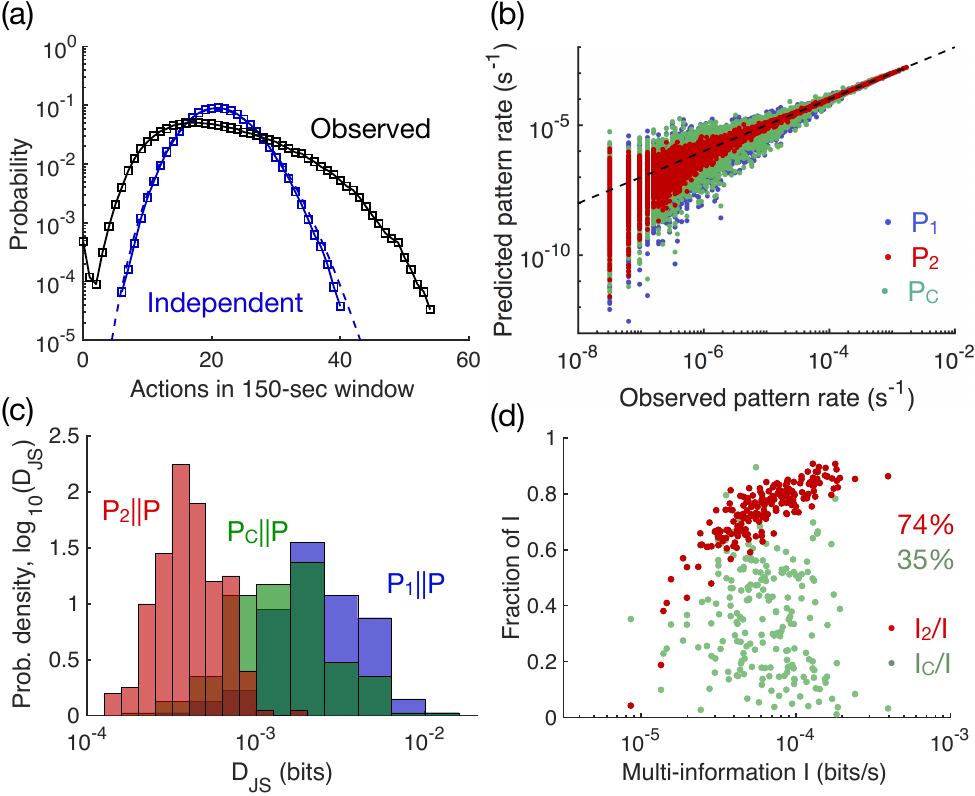}
\caption{\label{LastFM_fig} Performance of the maximum entropy model in a dataset of music streams. (a) Distribution of the number of streams in a given 150-second window in the dataset (black) and after shuffling individuals' activities to eliminate correlations (blue); dashed line indicates a Poisson fit to the shuffled data (blue). (b) The rate of each observed activity pattern across 200 randomly selected groups of 10 individuals is plotted against the approximate rates under the independent model $P_1$ (blue), the pairwise maximum entropy model $P_2$ (red), and the conditionally independent model $P_C$ (green); the dashed line indicates equality. (c) Jensen-Shannon divergences between the true distribution $P$ and the independent $P_1$ (blue), maximum entropy $P_2$ (red), and conditionally independent $P_C$ (green) models; the histograms reflect estimates from the 200 10-person groups. (d) Fraction of the network correlation (i.e., multi-information $I$) captured by the pairwise (red) and conditionally independent (green) models, plotted against the full multi-information; $I$ is divided by $\Delta t = $ 150 seconds to remove the dependence on window size.}
\end{figure}

To this point, all of our analysis has focused on modes of human activity that are themselves types of interactions between individuals. It is natural to suspect, therefore, that these activities might be particularly conducive to being described by a pairwise model. To test the ability of the pairwise maximum entropy model to describe other modes of human activity, here we consider a dataset of 610 individuals streaming music on the website \url{last.fm} over the span of one year \cite{Celma-01}. Discretizing the streaming activity into bins of width $\Delta t = $ 150 seconds (roughly corresponding to the length of an average song), we arrive at a set of $\sim 2\times 10^5$ activity vectors. Considering the number of music streams in a given 150-second window, we notice that the observed distribution is notably not described by an exponential distribution [Fig. \ref{LastFM_fig}(a)], which is attributable to the fact that the streaming data is much less sparse than any of the three activities studied previously. Nevertheless, we still find that the observed distribution is heavy-tailed relative to the independent Poisson distribution, and is characterized by surges of activity where upwards of 50 users are streaming music at a given time.

Randomly selecting 200 groups of 10 users, we show in Fig. \ref{LastFM_fig}(b) that the pairwise maximum entropy model provides a much tighter fit of the observed activity pattern rates than either the independent or conditionally independent models. Moreover, by studying the Jensen-Shannon divergences between the different models and the observed distribution of activity patterns, we find that we would need over six times as many data samples to distinguish $P_2$ from $P$ than to distinguish $P_1$ from $P$ and over four times more samples to distinguish $P_C$ [Fig. \ref{LastFM_fig}(c)]. These results are further supported by Fig. \ref{LastFM_fig}(d), which shows that the pairwise model captures $I_2/I \approx 74\%$ of the correlation in groups of 10 users, nearly identical to the case of face-to-face contacts. Meanwhile, the daily and weekly rhythms only account for $I_C/I_N \approx 35\%$ of the correlation in the data.

All together, our analysis of private messages, face-to-face contacts, and online music streams serve to strengthen the conclusions made in the main text; namely, that pairwise correlations can build upon one another to generate predictable patterns of population-wide activity.

\section{Learning a pairwise maximum entropy model: The inverse Ising problem}

\label{BM}

Here we present the theory and methodology behind learning a pairwise maximum entropy model of collective human activity. Specifically, we describe how to calculate the Ising parameters $h_i$ and $J_{ij}$ from a dataset of collective activity patterns. This inference task has a rich history in machine learning under the title Boltzmann machine learning \cite{Ackley-01} and is commonly referred to in physics as the inverse Ising problem \cite{Aurell-01}.

\begin{figure*}
\includegraphics[width = .8\textwidth]{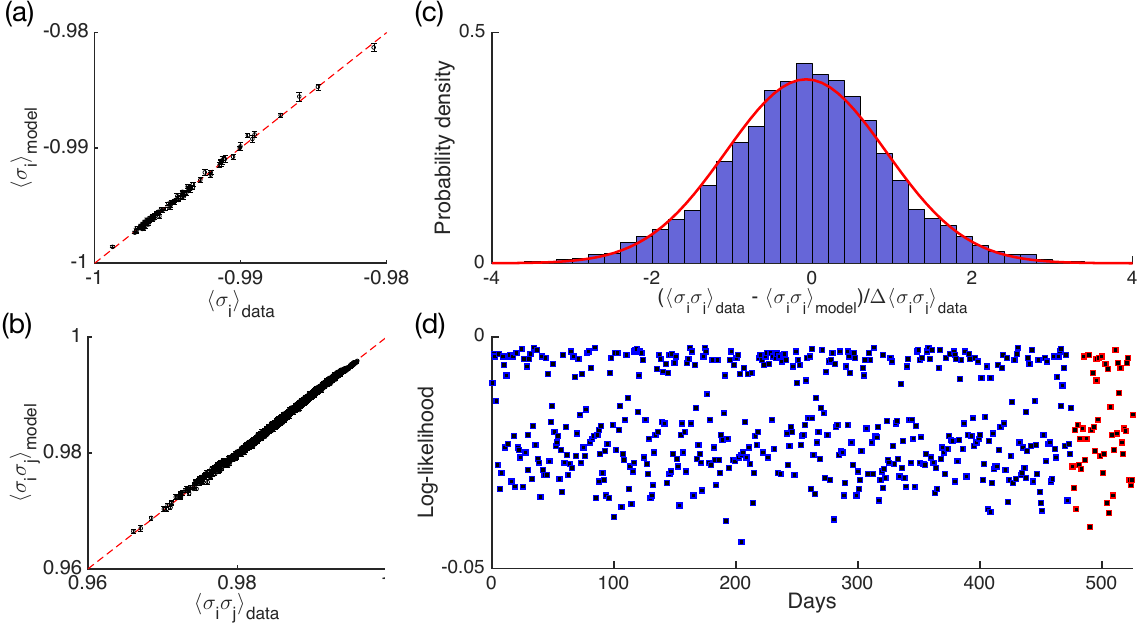}
\caption{\label{accuracy} Learning a pairwise maximum entropy model for a 100-person population. (a) Reconstructed activity rates for all 100 individuals under the maximum entropy model, plotted against their true activity rates. The dashed line indicates equality. (b) Reconstructed pairwise correlations under the maximum entropy model versus the observed correlations. (c) Distribution of the differences between the true and model pairwise correlations, normalized by the error in the data $\Delta\left<\sigma_i\sigma_j\right>$. For reference, the red line is a Gaussian distribution with unit variance. The empirically measured distribution has nearly Gaussian shape with standard deviation $\approx 1.05$, demonstrating that the learning algorithm reconstructs the pairwise correlations within experimental precision. (d) The per-person average log-likelihood of the data $\left<\log P_2(\bm{\sigma})\right>/N$, where the average is taken over all patterns within a given day, computed for the training days (blue) and test days (red). The data has been sorted so that the test days follow the training days, but the true choice of test days was random.}
\end{figure*}

\subsection{Exact models for small populations}

Given the observed distribution $P$ of activity patterns, there is a unique pairwise model $P_2$ that is consistent with the observed activity rates $\left<\sigma_i\right>$ and pairwise correlations $\left<\sigma_i\sigma_j\right>$, where $\left<\cdot\right>$ represents an average over $P$. To learn this pairwise model, one typically begins with an initial pairwise distribution $Q$ with parameters $\tilde{h}_i$ and $\tilde{J}_{ij}$ and then performs gradient descent in the model parameters, with gradients defined by
\begin{align}
\label{dh}
\Delta \tilde{h}_i &\propto \left<\sigma_i\right> - \left<\sigma_i\right>_Q, \\
\label{dJ}
\Delta \tilde{J}_{ij} &\propto \left<\sigma_i\sigma_j\right> - \left<\sigma_i\sigma_j\right>_Q,
\end{align}
where $\left<\cdot\right>_Q$ represents an average over $Q$. For groups of size $N=10$, these gradient calculations are tractable and standard gradient descent converges to the correct pairwise maximum entropy model $P_2$.

\subsection{Approximate models for large populations}

The primary difficulty in learning a maximum entropy model for a large population, such as the group of 100 email users, lies in calculating the one- and two-point correlations under $Q$ at each gradient step in Eqs. (\ref{dh}) and (\ref{dJ}). For large populations, exact calculations using the Boltzmann distribution are infeasible, and one must resort to approximate methods. The standard strategy is to simulate the system using Monte Carlo techniques \cite{Gilks-01,Ganmor-01,Tkacik-02}. Na\"{i}vely, one would run a new Monte Carlo simulation to estimate the gradients at each step of the learning algorithm. However, this straightforward approach is extremely inefficient. Instead, one can adjust the estimates of the one- and two-point correlations at each gradient step using importance sampling \cite{Jordan-01} or histogram Monte Carlo \cite{Ferrenberg-01}. In addition to limiting the number of Monte Carlo simulations, because each sample $\bm{\sigma}$ of $Q$ is dominated by inactive individuals, one can leverage sparse matrix operations to significantly speed up the simulations themselves.

We terminate the learning algorithm when the model correlations, $\left<\sigma_i\right>_Q$ and $\left<\sigma_i\sigma_j\right>_Q$, are sufficiently close to the observed correlations. The relevant scale for errors in the observed correlations is defined by the standard deviations $\Delta\left<\sigma_i\right>$ and $\Delta\left<\sigma_i\sigma_j\right>$, which are estimated by bootstrap sampling from the original dataset. Thus, the learning algorithm is terminated when 
\begin{align}
|\left<\sigma_i\right> - \left<\sigma_i\right>_Q| &< \Delta\left<\sigma_i\right> \approx 2.2\times 10^{-4} \\
|\left<\sigma_i\sigma_j\right> - \left<\sigma_i\sigma_j\right>_Q| &< \Delta\left<\sigma_i\sigma_j\right> \approx 1.7\times 10^{-4}.
\end{align}
We confirm that the individual email rates and pairwise correlations under the maximum entropy model $P_2$ match the observed correlations within the experimental errors in the data [Fig. \ref{accuracy}(a-c)].

For a population of 100 individuals, defining a pairwise maximum entropy model requires learning $N(N+1)/2=5050$ different parameters. Given such a large number, it is possible that the model is being finely tuned to match statistical errors in the data. To test for overfitting, we randomly select 476 of the 526 days to learn the model, and then we test the accuracy of the model on the remaining 50 days. We confirm that the pairwise model assigns the same amount of probability to the test data as to the training data, within errors, demonstrating that the learned model generalizes to describe data outside of the training set [Fig. \ref{accuracy}(d)]. We conclude that the learned pairwise model (i) fits the activity data within experimental precision and (ii) does not overfit statistical noise in the data. For access to the calculated external fields $h_i$ and pairwise interactions $J_{ij}$, please contact the corresponding author.

\section{The conditionally independent model}

\label{CondInd}

To test the prediction that collective behavior is driven by similarities in people's daily and weekly routines, we study the conditionally independent model $P_C$. Letting $p_i^t(\sigma_i)$ denote the probability of person $i$ performing an action within a window of width $\Delta t$ at time $t$ during the week, the conditionally independent model is defined by
\begin{equation}
P_C(\bm{\sigma}) = \frac{\Delta t}{\omega}\sum_t \prod_i p_i^t(\sigma_i),
\end{equation}
where $\omega$ denotes the length of a day or week. Under this conditionally independent model, correlations between individuals are driven by fluctuations in their inherent activity rates.

\section{Estimating entropy from finite data}

\label{Entropy}

To calculate the multi-information $I = S_1 - S$ of the network activity, we must compute the entropies of the independent model $S_1$ and the observed data $S$. While calculating $S_1$ is straightforward, we must estimate the true entropy $S$ from a finite number of samples, possibly leading to finite-size errors. Suppose that the dataset consists of the patterns $\{\bm{\sigma}^{\alpha}\}$ with corresponding probabilities $\{p^{\alpha}\}$. One could na\"{i}vely estimate the entropy using the standard formula
\begin{equation}
\tilde{S} = -\sum_{\alpha}p^{\alpha}\log p^{\alpha}.
\end{equation}
However, since some of the patterns are likely missing and the probabilities $p^{\alpha}$ are not exact, this estimate should fundamentally be viewed as an approximation to $S$ that improves as the number of samples increases. To correct for the sample size dependence of $\tilde{S}$ , we sub-sample the data and fit the resulting estimates using a form proposed by Strong et al. \cite{Strong-01},
\begin{equation}
\tilde{S}(\text{size}) = S + \frac{a}{\text{size}} + \frac{b}{\text{size}^2},
\end{equation}
where $a$ and $b$ are finite-size corrections. Using this fit, we can extract an accurate estimate of the true entropy $S$. We remark that for large datasets such as those considered here, and for relatively small networks like the 10-person groups studied in the main text, finite-size errors are small.

\section{Extended discussion}

\label{Discussion}

Our investigation of collective human behavior yields three distinct conclusions:
\begin{enumerate}
\item Large-scale behavior, characterized by surges in collective activity, cannot be understood using models that assume humans behave independently.
\item While collective behavior is far from independent, the minimal extension of the independent model consistent with the observed pairwise correlations captures most of the correlation in all populations considered, accurately predicting surges of collective activity.
\item In the network of email correspondence, the learned pairwise interactions are closely related to the underlying topology of inter-human communication, imbuing the maximum entropy model with real-world interpretability.
\end{enumerate}
Here we discuss the implications and limitations of these results, while keeping in mind that modern life involves a diverse range of activities, some of which may require a fundamentally different approach. Throughout, we emphasize important opportunities for future research.

\subsection{Internal correlations versus external influences}

In the study of human dynamics, as in the study of physical and biological systems, any macroscopic behavior that evades explanation by a model of independent elements fundamentally derives from two possible sources of correlation: (i) interactions between elements or groups of elements, and (ii) external influences on the system. In all human activities considered here, we witness surges of collective activity that cannot be explained under assumptions of human independence. Instead, we find that the populations are described quantitatively by models that include the simplest possible correlations---those between pairs of individuals. However, given that large-scale patterns could derive from higher-order correlations or from shared external inputs to the population, and given the myriad experiences that shape human actions, it would be na\"{i}ve to universally conclude that all collective human activity emerges from pairwise correlations. Instead, we hypothesize that particular activities fall along a spectrum, with internal correlations and external influences each playing roles of variable importance.

We remark that we have already witnessed evidence for such a spectrum in the different human activities studied above. For example, while patterns of email communication were reasonably well-described by taking into account people's weekly rhythms, capturing $\sim 67\%$ of the correlation structure in 10-person groups, private message correspondence had a markedly weak dependence on people's schedules, with daily routines accounting for only $\sim 5\%$ of the correlation in 10-person groups. These results agree with intuition, indicating that email activity is moderately tied to people's work and leisure schedules, while daily routines have nearly no predictive power in a network of private messages. Interestingly, correlations in both face-to-face contacts and online music streaming are moderately driven by daily and weekly routines, falling in between email and private message correspondence. With these results in mind, the clearest direction for future investigation is to continue probing different ends of the spectrum by quantifying the relative importance of internal correlations versus external influences in different modes of human behavior.

\subsection{The energy landscape of collective human behavior}

Every maximum entropy model $Q$ is defined by a Boltzmann distribution $Q(\bm{\sigma}) = \exp(-E(\bm{\sigma}))/Z$, where $E(\bm{\sigma})$ is the energy function, or Hamiltonian, that describes the system, and $Z$ is the normalization constant. In the case of the pairwise maximum entropy model, the relevant energy function is that of the Ising model, $E(\bm{\sigma}) = -\frac{1}{2}\sum_{i\neq j}J_{ij}\sigma_i\sigma_j - \sum_i h_i\sigma_i$. In statistical mechanics, there is a wealth of literature exploring the diversity of large-scale behaviors that can emerge from systems with different energy landscapes \cite{Chandler-01,Jaynes-01}. Thus, future research should leverage this connection to answer a number of important questions: What can the energy landscape of a given population tell us about its functional properties? Does collective human behavior favor dramatic shifts in activity, or are social populations organized to incentivize local fluctuations, guarding against the effects of large external shocks?

\subsection{Beyond equal-time correlations}

\label{Dynamic}

Throughout our analysis, we have focused on modeling equal-time correlations, which quantify the tendencies of individuals to engage in synchronous actions. In doing so, we have implicitly assumed that each observed activity pattern $\bm{\sigma}$ is drawn independently from an underlying stationary distribution $P(\bm{\sigma})$, leaving models of the population's activity without notions of time or causality. While studying equal-time correlations has allowed us to reach a number of important conclusions, the idea that patterns of human activity are sampled from a stationary distribution is not consistent with the common intuition that conscious human actions are often responses to prior social and environmental influences. For example, the fact that individuals perform bursts of actions in quick succession is thought to be the result of a decision-based queuing process \cite{Barabasi-01}, and it is known that the temporal scales of human activity can change over time \cite{Miritello-01,Borge-01,Navarro-01}.

In the context of human communication, a significant fraction of emails and private messages are direct responses to previous correspondence. Therefore, it would be interesting to study the correlations between people's activities with a time delay $\tau$ in between, where $\tau$ represents the characteristic response time of communication in the population. Such spatiotemporal correlations have recently received a large amount of interest in neuroscience and biology, where it has been found that the spatiotemporal patterns of spiking neurons in the brain and flocks of birds in flight are only partially captured by stationary maximum entropy models \cite{Tang-01,Marre-01,Cavagna-01}. Similarly, studying the spatiotemporal patterns that define collective human activity has significant implications for understanding the causal flow of influences and information between individuals in a population \cite{Navarro-01}. Furthermore, developing accurate dynamical models of large-scale behavior has important ramifications for predicting the effects of interventions and time-varying perturbations in networks of interacting humans \cite{Candia-01,Crane-01,Peng-01,Onnela-01,Gonzalez-01,Caldarelli-01,Pastor-01}.

\bibliography{BMSocialNetworkBib}

\end{document}